\documentclass[fleqn,usenatbib]{mnras}

\usepackage{newtxtext,newtxmath}
\usepackage{xcolor}
\usepackage[T1]{fontenc}

\DeclareRobustCommand{\VAN}[3]{#2}
\let\VANthebibliography\thebibliography
\def\thebibliography{\DeclareRobustCommand{\VAN}[3]{##3}\VANthebibliography}

\usepackage{graphicx}	% Including figure files
\usepackage{amsmath}	% Advanced maths commands

\usepackage{enumitem}

\title[Hi-Fi reproduction of central galaxy properties]{High-fidelity reproduction of central galaxy joint distributions with Neural Networks}

\author[N.V.N. Rodrigues et al.]{
Nat\'alia V. N. Rodrigues,$^{1}$\thanks{E-mail: natalia.villa.rodrigues@usp.br}
Natal\'i S. M. de Santi,$^{1, 2}$
Antonio D. Montero-Dorta,$^{3}$
L. Raul Abramo$^{1}$
\\
$^{1}$Departamento de F\'isica Matem\'atica, Instituto de F\'{\i}sica, Universidade de S\~ao Paulo, Rua do Mat\~ao 1371, CEP 05508-090, S\~ao Paulo, Brazil\\
$^{2}$Center for Computational Astrophysics, Flatiron Institute, 162 5th Avenue, New York, NY, 10010, USA\\
$^{3}$  Departamento de F\'isica, Universidad T\'ecnica Federico Santa Mar\'ia, Casilla 110-V, Avda. Espa\~na 1680, Valpara\'iso, Chile\\
}

\date{Accepted XXX. Received YYY; in original form ZZZ}

\pubyear{2023}

\begin{document}
\label{firstpage}
\pagerange{\pageref{firstpage}--\pageref{lastpage}}
\maketitle

\begin{abstract}
The relationship between galaxies and haloes is central to the description of galaxy formation, and a fundamental step towards extracting precise cosmological information from galaxy maps. 
However, this connection involves several complex processes that are interconnected. 
Machine Learning methods are flexible tools that can learn complex correlations between a large number of features, but are traditionally designed as deterministic estimators.
In this work, we use the IllustrisTNG300-1 simulation and apply neural networks in a binning classification scheme to predict probability distributions of central galaxy properties, namely stellar mass, colour, specific star formation rate, and radius, using as input features the halo mass, concentration, spin, age, and the overdensity on a scale of 3 $h^{-1}$ Mpc.
% We verify that the predictions are consistent with our understanding of the halo--galaxy connection by analysing the predicted distributions for individual haloes. 
The model captures the intrinsic scatter in the relation between halo and galaxy properties, and can thus be used to quantify the uncertainties related to the stochasticity of the galaxy properties with respect to the halo properties.
In particular, with our proposed method, one can define and accurately reproduce the properties of the different galaxy populations in great detail. 
We demonstrate the power of this tool by directly comparing traditional single-point estimators and the predicted joint probability distributions, and also by computing the power spectrum of a large number of tracers defined on the basis of the predicted colour-stellar mass diagram. We show that the neural networks reproduce clustering statistics of the individual galaxy populations with excellent precision and accuracy.
\end{abstract}

\begin{keywords}
galaxies: statistics -- cosmology: large-scale structure of Universe -- methods: data analysis -- methods: statistical
\end{keywords}

%%%%%%%%%%%%%%%%%%%%%%%%%%%%%%%%%%%%%%%%%%%%%%%%%%

%%%%%%%%%%%%%%%%% BODY OF PAPER %%%%%%%%%%%%%%%%%%

\section{Introduction}

Characterising the connection between the properties of galaxies and those of the underlying population of dark-matter (DM) haloes is one of the most crucial aspects to understand the large-scale structure (LSS) of the Universe.
This link not only encapsulates fundamental information about the process of galaxy formation, but it is also a crucial step to optimise the extraction of cosmological constraints from galaxy maps.

The halo--galaxy connection is nowadays investigated using a variety of techniques (see, e.g., \citealt{Wechsler2018}). 
On the one hand, empirical methods use DM-only simulations as the basis on top of which different analytical prescriptions are implemented in order to establish that connection. 
These techniques include sub-halo abundance matching (SHAM, e.g., \citealt{Conroy2006,Behroozi2010,Trujillo-Gomez2011,Favole2016,Guo2016,Contreras2020,Contreras2021,Hadzhiyska2021, Favole2021}), halo occupation distributions (HODs, e.g., \citealt{Berlind2002, Zehavi2005,Zehavi2018,Artale2018,Bose2019,Hadzhiyska2020B,Xu2021}) and empirical forward modelling (e.g., \citealt{becker2015_eam, moster2018_emerge, behroozi2019_um}). 
On the other hand, it is possible to model, with varying degrees of detail, the physical mechanisms that shape the process of galaxy formation. 
In this context,  hydrodynamical simulations (e.g., \citealt{Somerville2015,Naab2017,Pillepich2018b,Pillepich2018, Springel2018, camels_presentation, camels_data_release})
are perhaps the most ambitious efforts. 
These models employ known physics to simulate, at a sub-grid level, a variety of processes that are related to galaxy formation such as star formation, radiative metal cooling, and supernova, stellar, and black hole feedback -- for reviews on this, see \citealt{Somerville2015,Naab2017}. 
This modelling can also be approached from a semi-analytic, less computationally demanding, perspective. 
These semi-analytic models (SAMs, e.g., \citealt{White1991, Guo2013}) employ physically motivated recipes to mimic the galaxy formation processes.

In this paper, we investigate the halo--galaxy connection from a machine learning (ML) perspective.
The issue of the halo--galaxy connection has been addressed using ML by many works (e.g., \citealt{Kamdar2016, Agarwal2018, Calderon2019, Jo2019, Man2019,Yip2019, Zhang2019,Jo2019,Kasmanoff2020, Delgado2021,McGibbon2021,shao2021,Lovell2021, hg_scatter_2022, de_Andres_2022, Jespersen_2022, 2023MNRAS.518.5670C}).
In \citet{deSanti2022} we provide a ML suite combining some of the most powerful, well-known models in the literature to predict central galaxy properties using host halo properties. 
All the applied methods, however, are designed to return a single value for each galaxy property, 
independently of the remaining properties.
However, there are many complex interrelated processes involved in the formation and evolution of galaxies, and their properties cannot be precisely determined by halo properties alone. 
Therefore, a model that proposes to map the relation between galaxies and host haloes should encode not only the correlations between galaxy properties, but also the uncertainties due to the stochastic aspects of galaxy formation.
In other words, any given halo could host a central galaxy with a variety of properties and, hence, a model should return joint probability distributions for the possible values of those galaxy properties, instead of a single one.

The ML suite from our precursor work \citep{deSanti2022} provided encouraging results in terms of single-point estimation metrics, such as the Pearson correlation coefficient between true and predicted values, especially for stellar mass, which is highly correlated with halo mass. 
However, deterministic models that try to predict individual galaxy properties can be biased towards the most frequent values, and thus fail to recover the overall distributions of the galaxy properties.
In that paper, this issue is treated as an imbalanced data problem, i.e., despite of the fact that different output values could be associated with some fixed set of halo properties, the machine tends to assign the most frequent values. 
To address this problem, we made use of a data augmentation technique to increase the weight of the less represented instances, which allowed us to better recover the under-represented populations, but still in a way that each halo is assigned a single, individual value for each central galaxy property \citep{deSanti2022}.

In the present work, we proceed by predicting probability distributions with neural networks (NNs) with a binning classification scheme, which we refer to as $\rm NN_{class}$, for the same central galaxy properties as \cite{deSanti2022}, namely, stellar mass, $g - i$ colour, specific star formation rate, and galaxy radius.
This not only enables us to recover the overall distributions of the galaxy properties from the IllustrisTNG300-1 (hereafter, TNG300) sample, but also to capture the intrinsic scatter in the halo--galaxy mapping by providing, for each halo, the probability distributions associated with its central galaxy properties. We also train $\rm NN_{class}$ to predict the galaxy properties jointly, finding that the joint distributions recover correlations that are lost when predicting univariate distributions independently.
ML probability-based descriptions have been used in related contexts, in particular with NNs, such as photometric redshift estimation (e.g., \citealt{Lima_2022}), dynamical mass of galaxy clusters estimation (e.g., \citealt{Ho_2021, Kodi_Ramanah_2020}) and recently in the halo--galaxy connection (e.g., \citealt{hg_scatter_2022}).

In order to study how $\rm NN_{class}$ captures the intrinsic stochasticity in the halo--galaxy connection, we analyse the shape of the distributions of individual galaxies, which gives some insights on the contribution of secondary halo properties.
Moreover, we analyse how this uncertainty affects clustering statistics, namely the power spectrum. 
Our technique enables us to define as many galaxy populations as wished, and to analyse to what extent those populations occupy the same types of haloes. 
We explore this flexibility by computing the power spectrum of a large number of galaxy populations (tracers), selected on the basis of the colour-stellar mass diagram.

The paper is organised as follows. The IllustrisTNG data and the chosen set of halo and galaxy properties are described in \S\ref{data}. 
In \S\ref{methodology}, we explain how we applied NNs to predict joint probability distributions.
Section \ref{results} analyses the quality of the results obtained with the NNs by comparing the predictions with the IllustrisTNG catalogue. 
In \S\ref{power spectrum}, we present our results in terms of the power spectra of several galaxy populations. 
Finally, we outline our main conclusions in  \S\ref{conclusions}, and discuss our plans for future improvements and applications.

\section{Data}\label{data}

Our analysis is based on data from the IllustrisTNG magnetohydrodynamical cosmological simulation \citep{Pillepich2018b,Pillepich2018,Nelson2018_ColorBim,Marinacci2018,Naiman2018,Springel2018,Nelson2019}. 
This simulation suite, which was generated using the {\sc arepo} moving-mesh code \citep{Springel2010}, is an improved version of the previous Illustris simulation \citep{Vogelsberger2014a, Vogelsberger2014b, Genel2014}. 
IllustrisTNG features a variety of updated sub-grid models accounting for star formation, radiative metal cooling, chemical enrichment from SNII, SNIa, and AGB stars, as well as feedback mechanisms (including stellar and super-massive black hole feedback). 
These models were calibrated to reproduce an array of observational constraints, such as the $z=0$ galaxy stellar mass function and the cosmic SFR density, to name but a few (see the aforementioned references for more information). 
The IllustrisTNG simulation adopts the standard $\Lambda$CDM cosmology \citep{Planck2016}, with parameters $\Omega_{\rm m} = 0.3089$,  $\Omega_{\rm b} = 0.0486$, $\Omega_\Lambda = 0.6911$, $H_0 = 100\,h\, {\rm km\, s^{-1}Mpc^{-1}}$ with $h=0.6774$, $\sigma_8 = 0.8159$, and $n_s = 0.9667$.

The ML methodology that we developed in this work to reproduce the halo--galaxy connection is applied to galaxy clustering in terms of the power spectrum. 
For this reason, in order to minimise cosmic variance, we chose to analyse the largest box available in the database, TNG300, spanning a side length of $205\,\,h^{-1}$Mpc with periodic boundary conditions. 
TNG300 contains 2500$^3$ DM particles of mass $4.0 \times 10^7$ $h^{-1} {\rm M_{\odot}}$ and 2500$^3$ gas cells of mass $7.6 \times 10^6$ $h^{-1} {\rm M_{\odot}}$. 
The adequacy of TNG300 in the context of clustering science has been extensively proven in a variety of analyses (see, e.g., \citealt{Contreras2020,Gu2020,Hadzhiyska2020,MonteroDorta2020,Shi2020,Hadzhiyska2021,MonteroDorta2021A,MonteroDorta2021B,Favole2021,deSanti2022}). 

In this work, we employ both galaxy and DM halo information from TNG300. 
DM haloes in the entire IllustrisTNG suite are identified using a friends-of-friends (FOF) algorithm based on a linking length of 0.2 times the mean of the inter-particle separation \citep{Davis1985}. 
As in \cite{deSanti2022}, the following halo properties are used as input features to train the NNs:

\begin{itemize}[leftmargin=*]

    \item {\it{Virial mass}} ($M_{\rm vir} [h^{-1} {\rm M_{\odot}}]$), which is computed by adding up the mass of all gas cells and particles contained within the virial radius $R_{\rm vir}$ (based on a collapse density threshold of $\Lambda_c =200$). In order to ensure that haloes are well resolved, we impose a mass cut $\log_{10}(M_{\rm vir}[h^{-1}{\rm M_{\odot}}]) \geq 10.5$, corresponding to at least 500 dark matter particles.
    
    \vspace{0.1cm}
    \item {\it{Virial concentration}} ($c_{\rm vir}$), defined in the standard way as the ratio between the virial radius and the scale radius, i.e.,  $c_{\rm vir} = R_{\rm vir}/R_{\rm s}$. $R_{\rm s}$ is obtained by fitting the DM density profiles of individual haloes with a NFW profile \citep{nfw1997}.

    \vspace{0.1cm}
    \item {\it{Halo spin}} ($\lambda_{\rm halo}$), for which we follow the \cite{Bullock2001_2} definition: $\lambda_{\rm halo} = |J|/\sqrt{2} { M_{\rm vir}} {V_{\rm vir}} { R_{\rm vir}}$. Here, $J$ and $V_{\rm vir}$ are the angular momentum of the halo and its circular velocity at $R_{\rm vir}$, respectively.

    \vspace{0.1cm}
    \item {\it{Halo age}}, parametrised as the half-mass formation redshift $z_\text{1/2}$. This parameter corresponds to the redshift at which half of the present-day halo mass has been accreted into a single subhalo for the first time. The formation redshift is measured following the progenitors of the main branch of the subhalo merger tree computed with {\sc sublink}, which is initialised at $z = 6$. 
	
	\vspace{0.1cm}
    \item The {\it{overdensity}} around haloes on a scale of 3 $h^{-1}$Mpc ($\delta_3$), defined as the number density of subhaloes within a sphere of radius $R = 3 h^{-1}$Mpc, normalised by the total number density of subhaloes in the TNG300 box \citep[e.g.,][]{Artale2018,Bose2019}.
    
\end{itemize}

On the other hand, subhaloes (i.e., gravitationally bound substructures) are identified in IllustrisTNG using the {\sc subfind} algorithm \citep{Springel2001,Dolag2009}. Subhaloes containing a non-zero stellar mass component are labelled as galaxies. Again, following \cite{deSanti2022} for consistency, TNG300 galaxies are characterised in this work using the following basic properties:

\begin{itemize}[leftmargin=*]
    \item The {\it{stellar mass}} ($M_\ast{}$ [$h^{-1} {\rm M_{\odot}}$]), which includes all stellar particles within the subhalo. In order to ensure that galaxies are well resolved, we impose a mass cut $\log_{10}(M_{\ast{}}[h^{-1} {\rm M_{\odot}}]) \geq 8.75$, corresponding to at least 50 gas cells.
    %Note that this cut has little effect on the performance of our method.
    \vspace{0.1cm}
    \item The {\it{colour}} $g - i$, computed from the rest-frame magnitudes, which are obtained in IllustrisTNG by adding up the luminosities of all stellar particles in the subhalo (\citealt{Buser1978}). Note that the specific choice of colour is rather arbitrary. We have checked that using other combinations (i.e., $g - r$) provides similar results. 

    \vspace{0.1cm}
    \item The {\it{specific star formation rate}} (sSFR [$ {\rm yr^{-1}} h$]), which is the  star formation rate (SFR) normalised by stellar mass. The SFR is computed by adding up the star formation rates of all gas cells in the subhalo. Note that around $14\%$ of the galaxies at redshift $z=0$ in TNG300 have SFR$=0$. In order to avoid numerical issues, we have adopted the same approach as in \cite{deSanti2022}, assigning to these objects artificial values of SFR sampled from a Gaussian distribution $\mathcal{N} (\mu = - 13.5, \sigma = 0.5)$.
    %The sSFR in our sample covers sSFR $\in [-17.00, -8.30] \rm yr^{-1} M_{\odot}$.
    \vspace{0.1cm}
    \item The {\it{galaxy size}}, parameterised as the stellar (3D) half-mass radius ($R_{1/2}^{(*)} [h^{-1} \, {\rm kpc}]$) -- i.e., %$R_{1/2}^{(*)}$ is 
    the comoving radius containing half of the stellar mass in the subhalo.
    %The galaxy radius in our sample covers $R_{1/2}^{(*)} \in [0.05, 2.00] h^{-1} \, {\rm kpc}$.
\end{itemize}

\section{Methodology}\label{methodology}

NNs are designed to learn how to map an instance, which is characterised by some set of input features $X$, to a set of output features $Y$, by weighting and combining the input features. 
These weights are fitted by minimising a loss function with some optimiser.

In this work, the input features are the halo properties and the outputs are the galaxy properties introduced in \S\ref{data}. 
Starting with a sample where the target value $Y$ is known for all instances (the TNG300 catalogue), we split it into training, validation and test sets. 
The training set is used to fit the model parameters (weights). 
The validation set is used to monitor overfitting, i.e., to ensure that the model is properly generalising to data outside of the training set, and to fit the model's hyperparameters\footnote{In a NN, the model's parameters are the weights to be learned automatically, while the hyperparameters are the number of layers, neurons, number of epochs, etc., which are often chosen manually.}.
The test set remains completely blind to the training and validating procedures, and can thus be used to infer the performance of the model when applied to entirely new instances.
The training, validation and test sets contain, respectively, 48\%, 12\% and 40\% of the initial sample of 174,527 objects from the TNG300 catalogue.

Our goal is to predict central galaxy properties from a set of halo properties. 
In the context of ML, this would in principle fall in the category of a supervised regression problem. 
However, traditional regression models are designed to output single values, while any given halo could host many different central galaxies (since the set of halo properties that we use as inputs do not determine exactly the outcome of the galaxy formation process in terms of the precise values of the galaxy properties).
This is reflected, as an example, in the well-known scatter in the stellar-to-halo mass relation (\citealt{Wechsler2018, hg_scatter_2022}). %{\color{blue} [Raul: citation here perhaps?]}
Therefore, in order to incorporate this uncertainty, we need a model that returns not only a single best-estimate value for each galaxy property, but some proxy for the probability distribution for those properties.

In this paper, we have addressed this issue
by converting the regression problem into a classification. 
The idea is to define $K$ classes by splitting each galaxy property into $K$ intervals, or bins.
Just like in the usual classification tasks, the model will return a score associated with each class (bin). 
These scores add up to one, giving a probabilistic interpretation of the output. This approach has been widely used, as an example, in the context of photometric redshift estimation \citep{ANNz_2016, Pasquet_2018, Lima_2022}.
We refer to our method, which is based on training NNs classifiers, as $\rm NN_{class}$.

As a starting point, we train four models to predict each galaxy property individually as univariate distributions, i.e., we have separate models to predict $P(M_*), P(g - i), P(\text{sSFR}), P(R_{1/2}^{(*)})$. 
As we discuss in \S \ref{results}, this approach is sufficient to recover the overall distribution $P(Y)$ for a given sample. 
However, this does not guarantee, \textit{a priori}, that the joint distributions are well reproduced.
Therefore, we proceed to predict jointly pairs of properties, namely $P(M_*, g-i), P(M_*, \text{sSFR}), P(g-i, \text{sSFR})$ and $ P(R_{1/2}^{(*)}, M_*)$. 
Our strategy is similar to the univariate $P(Y)$ case: we make a grid in the $\{ Y_1, Y_2\}$ subspace in such a way that the output corresponds to pixels in this grid.
Although in this paper we restrict ourselves to only two galaxy properties when predicting joint distributions, a similar approach could be used, in principle, to characterise galaxies and define populations using an arbitrary number of properties. 
This generalisation will be implemented in an upcoming paper.

Unless otherwise stated, for all the results shown here we set $K = 50$ classes for each one of the central galaxy properties, in equally spaced bins. 
For stellar mass, for example, this corresponds to bins of $0.085$ dex. 
We must draw attention to the fact that this choice of binning is arbitrary. 
We have tried different numbers of bins, finding similar results in terms of the recovery of the distributions. Note that 
more refined versions of NNs that output distributions without binning the properties, and thus keeping it as a regression problem, already exist in the literature.
In the context of photo-z estimation, \cite{Lima_2022}, for example, compares different types of NNs that return distributions, such as Mixture Density Networks \citep{Bishop1994MixtureDN},  Bayesian NNs, and also NNs following a similar strategy as in this work, with a binning classification scheme.
\cite{Ho_2021} estimate the probability distribution of the dynamical mass of galaxy clusters and also compare several types of NNs, including a classifier which is similar to our $\rm NN_{class}$.
In the context of the halo--galaxy connection, \cite{hg_scatter_2022} model the stellar-to-halo mass relation scatter with a Gaussian distribution and train an ensemble of NNs that predicts the mean and standard deviation.
We found the binned classification to be a simpler approach that works as a proof of concept. 
A more careful exploration of alternative methods is left as future refinements.

Throughout the analysis, we compare our $\rm NN_{class}$ method with the deterministic models developed by \cite{deSanti2022}, which we use as our baseline. 
In that work, several ML models are combined to return a final, consensus output for the same galaxy properties described in \S\ref{data}. 
The two consensus estimators are built from either the ``Raw'' models, which were trained with the original TNG300 sample, or the ``SMOGN'' models, which were trained using a data-augmented version of that data set.
The SMOGN models were developed because of the difficulty for Raw models to recover the least frequent values of galaxy properties -- i.e., to reproduce the tails of the distributions. 
The SMOGN data augmentation technique is a strategy to handle imbalanced data sets, whereby additional objects are artificially introduced in the training sample in order to force the machine to give more importance to less represented objects \citep{Kunz2019}. 

The specifications of $\rm NN_{class}$ are described as follows. 
We use the categorical cross-entropy loss function and the \texttt{adam} optimiser to train the networks. 
The architecture may change depending on the galaxy properties to be predicted. 
In general, our developed networks have a single intermediate layer, with a number of neurons that typically depends on whether the output is an univariate or a joint distribution. 
We use the L2 regularisation, which applies a penalty proportional to the square of the model's weights. 
The number of epochs (iterations) is constrained with an early-stopping criteria based on the validation set loss. 
In the intermediate layers we used the ReLU function as activation, while in the output layer we use the Softmax function, which is similar to the Sigmoid function, but it normalises the output in such a way that the scores of the $K$ classes add up to one. 
In this way, the $\rm NN_{class}$ output works as a proxy for a probability in bins of galaxy properties.

\section{Results}\label{results}

\begin{figure*}
    \includegraphics[scale=0.15]{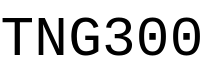}
    \hspace{3cm}
    \includegraphics[scale=0.15]{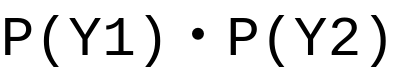}
    \hspace{3cm}
    \includegraphics[scale=0.15]{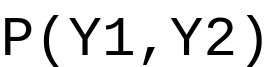}\\
    \centering
    \includegraphics[scale=0.225]{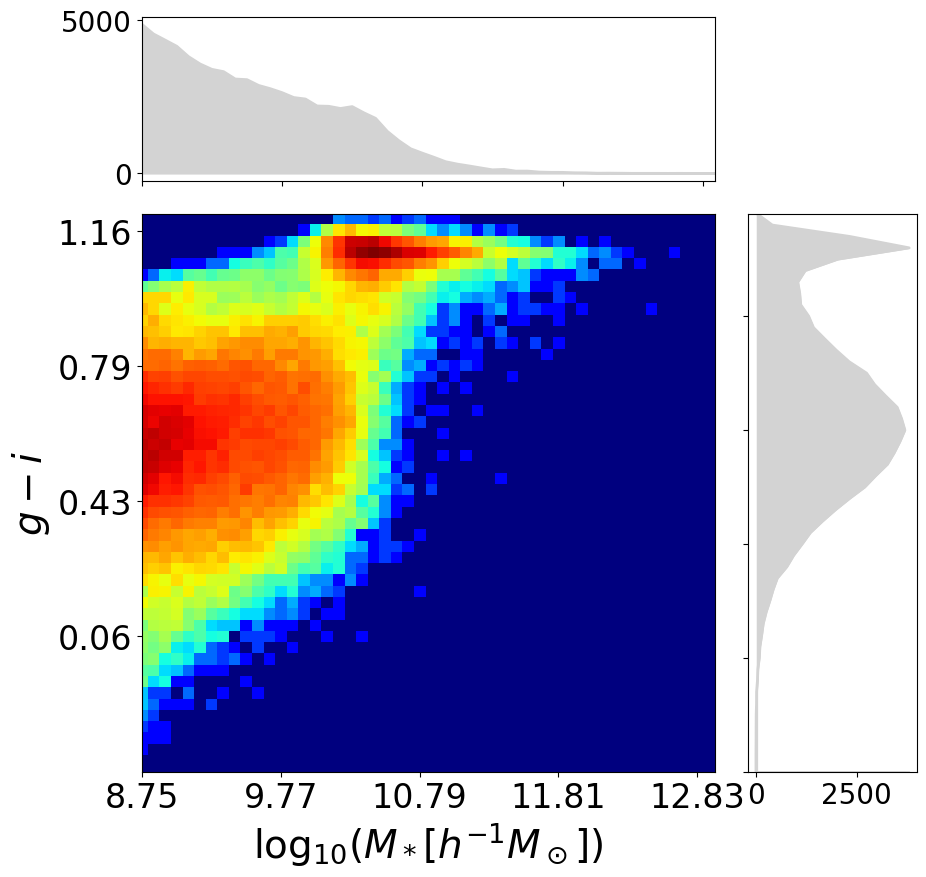}
    \hspace{0.3cm}
    \includegraphics[scale=0.225]{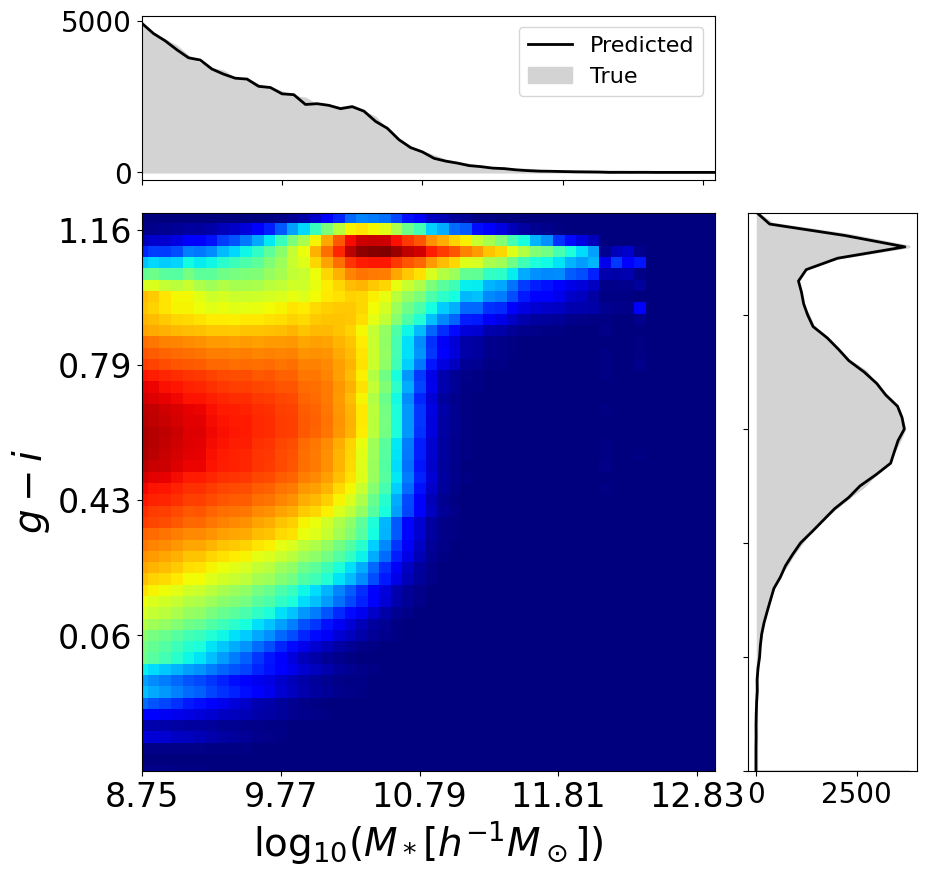}
    \hspace{0.3cm}
    \includegraphics[scale=0.225]{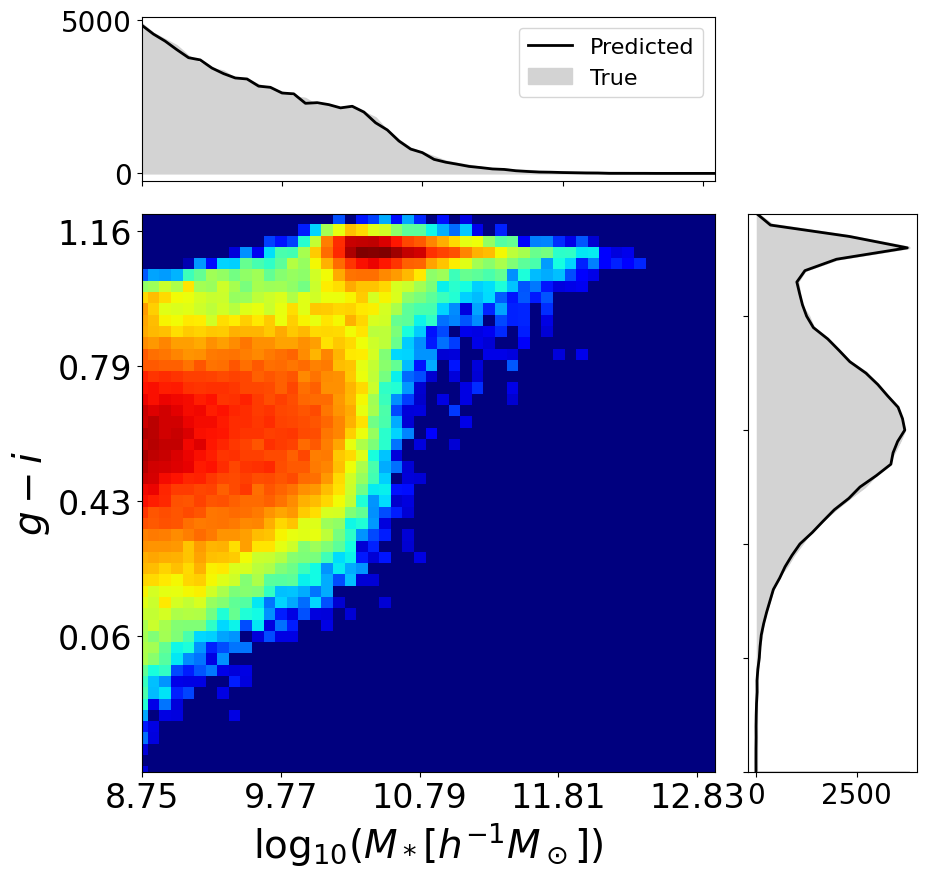}\\
    \vspace{0.2cm}
    \includegraphics[scale=0.225]{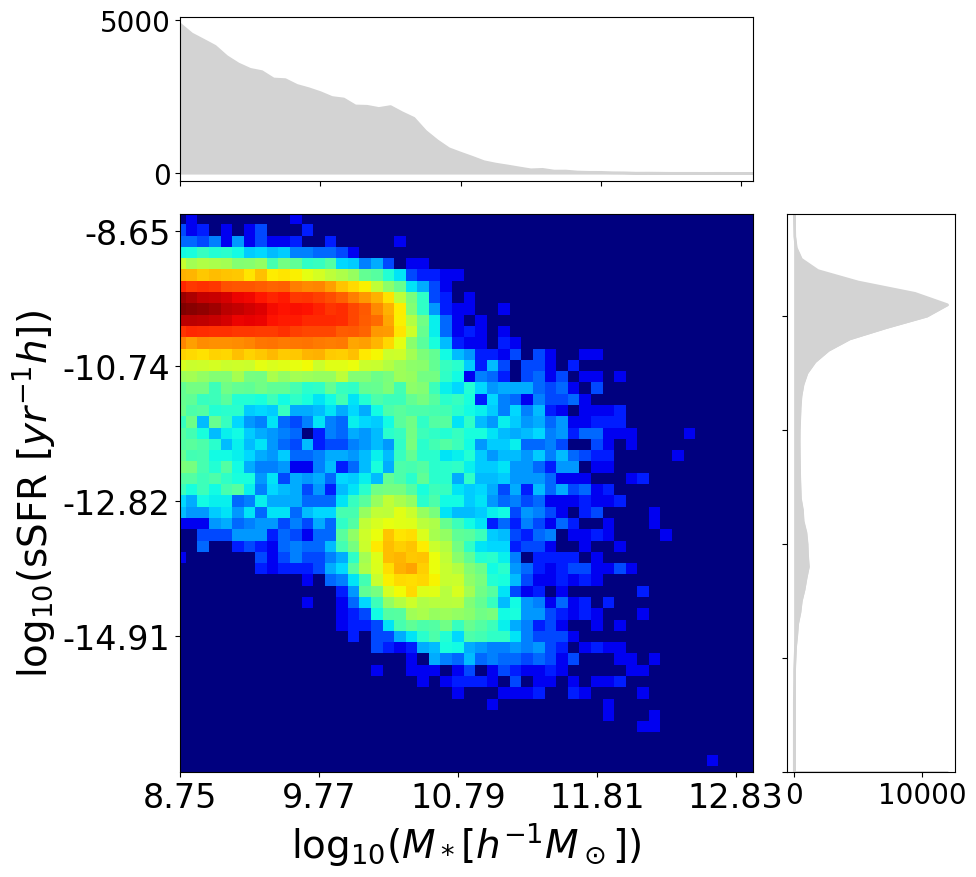}
    \includegraphics[scale=0.225]{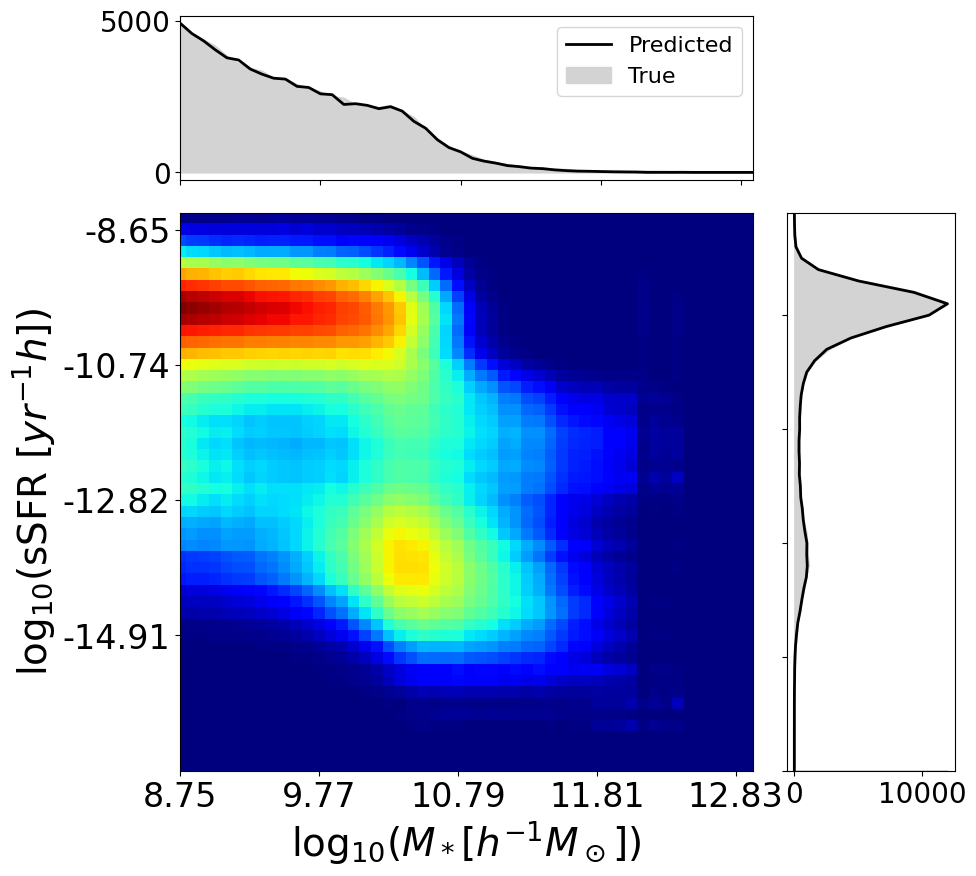}
    \includegraphics[scale=0.225]{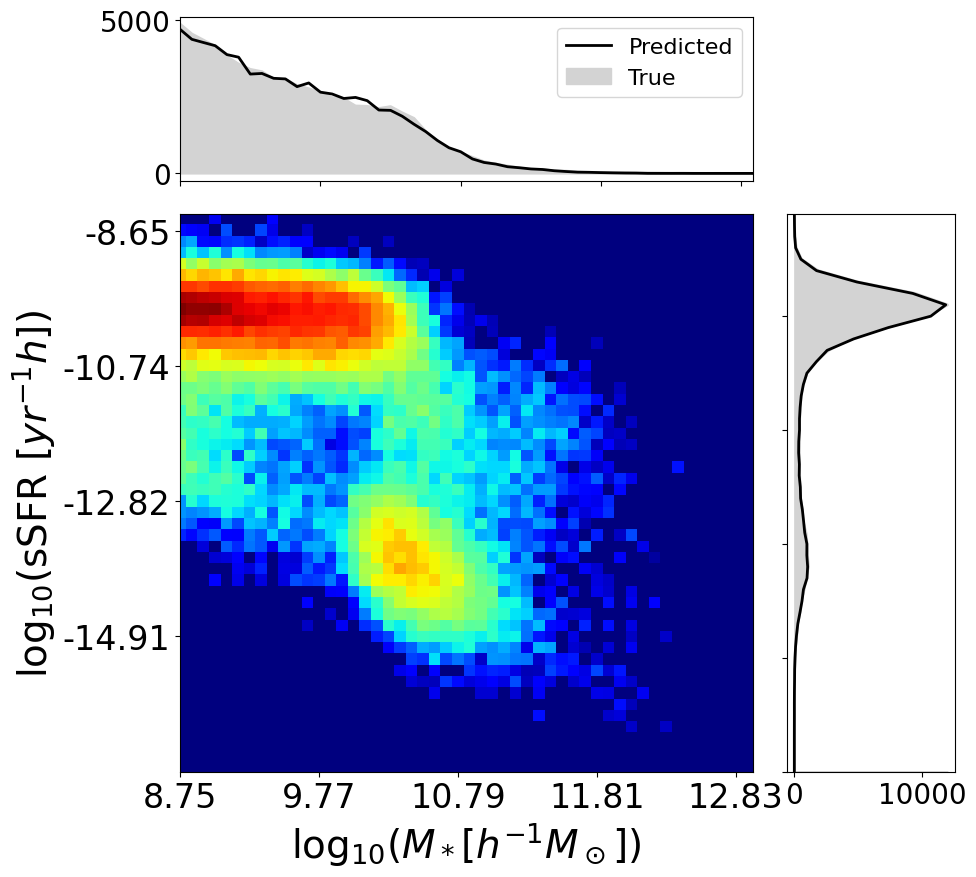}\\
    \vspace{0.2cm}
    \includegraphics[scale=0.225]{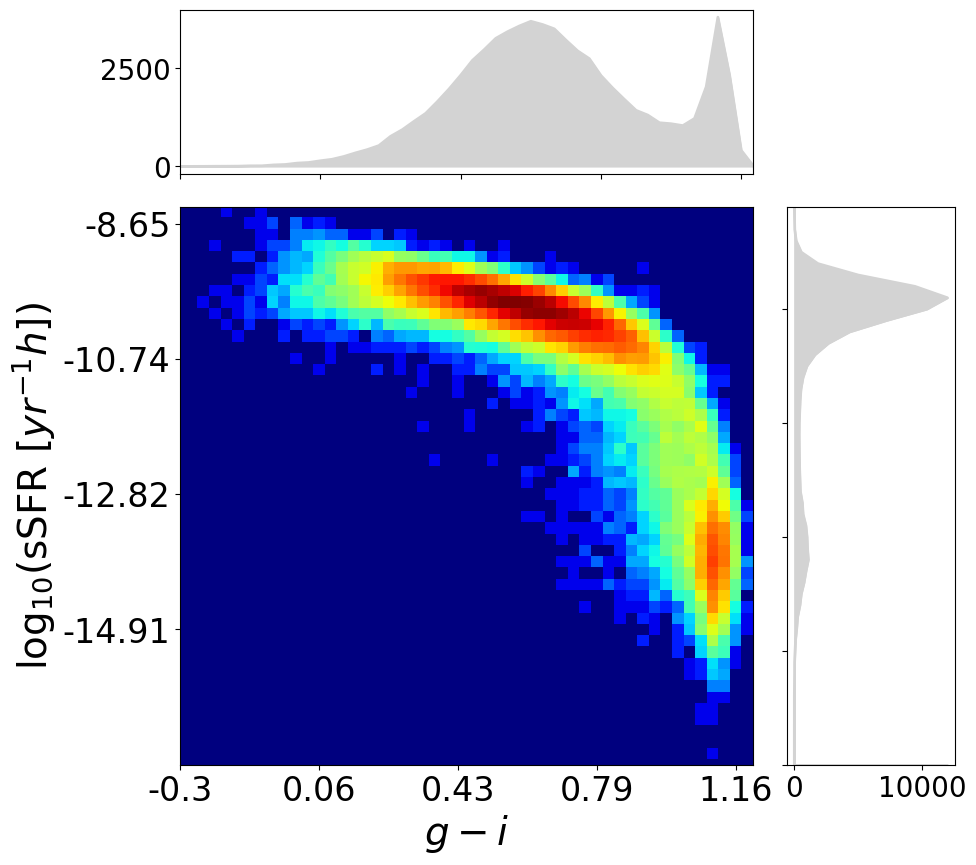}
    \includegraphics[scale=0.225]{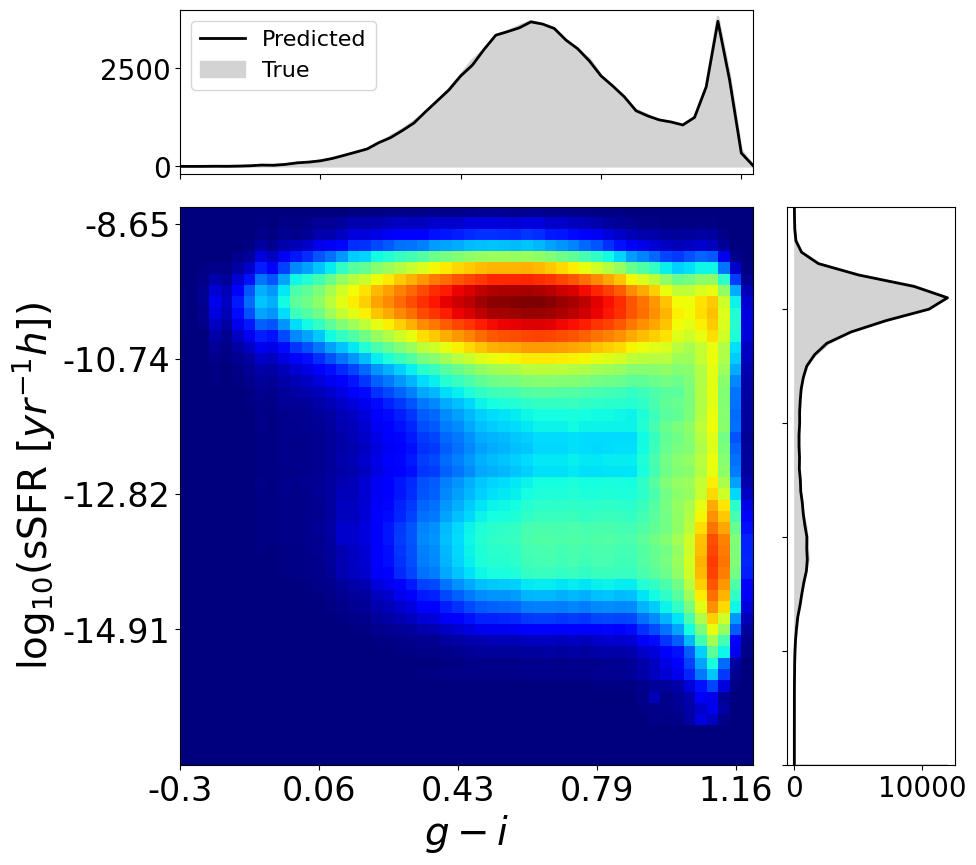}
    \includegraphics[scale=0.225]{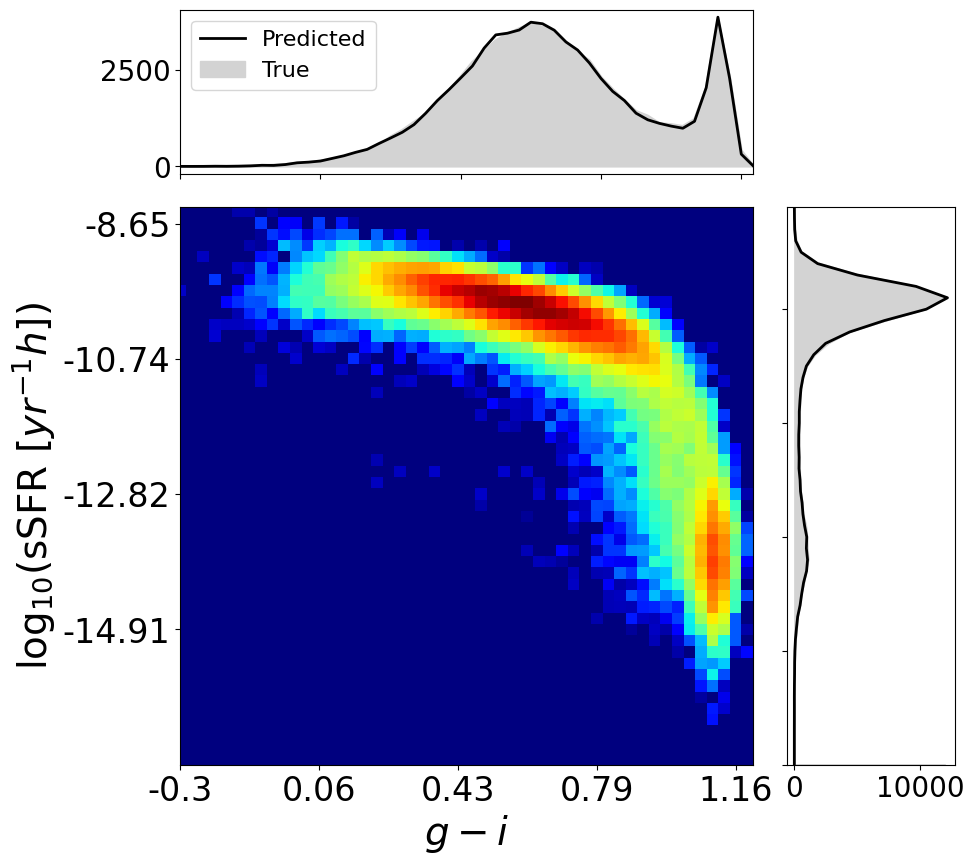}\\
    \vspace{0.2cm}
    \includegraphics[scale=0.225]{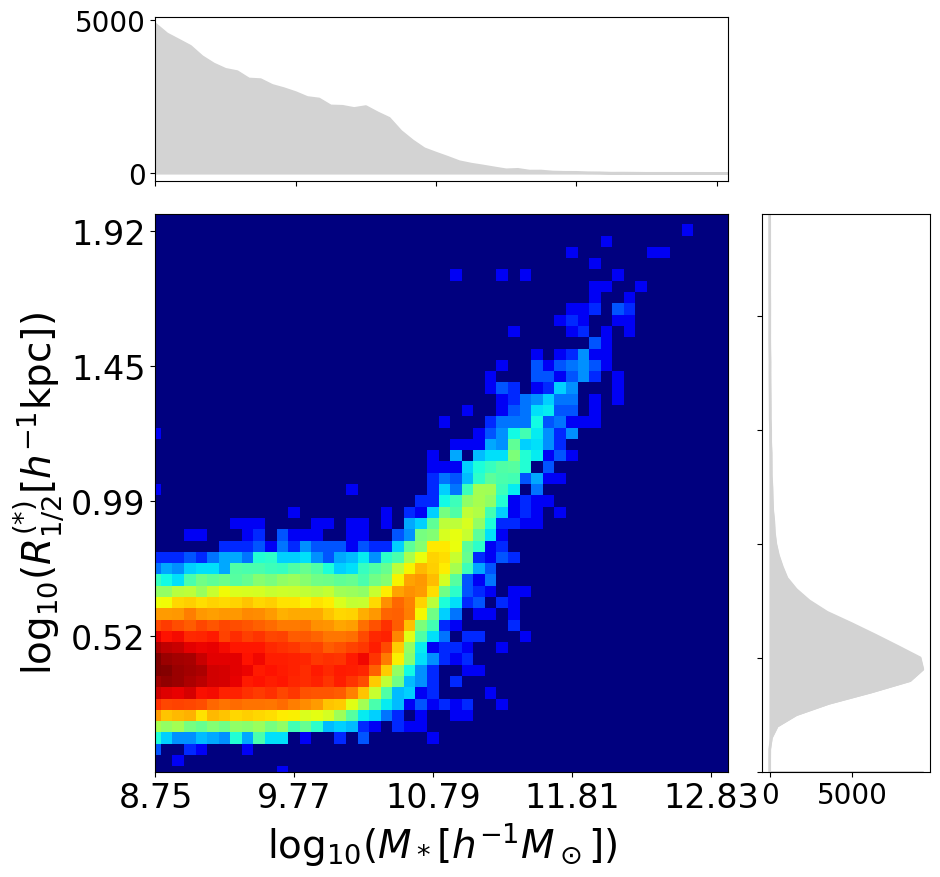}
    \includegraphics[scale=0.225]{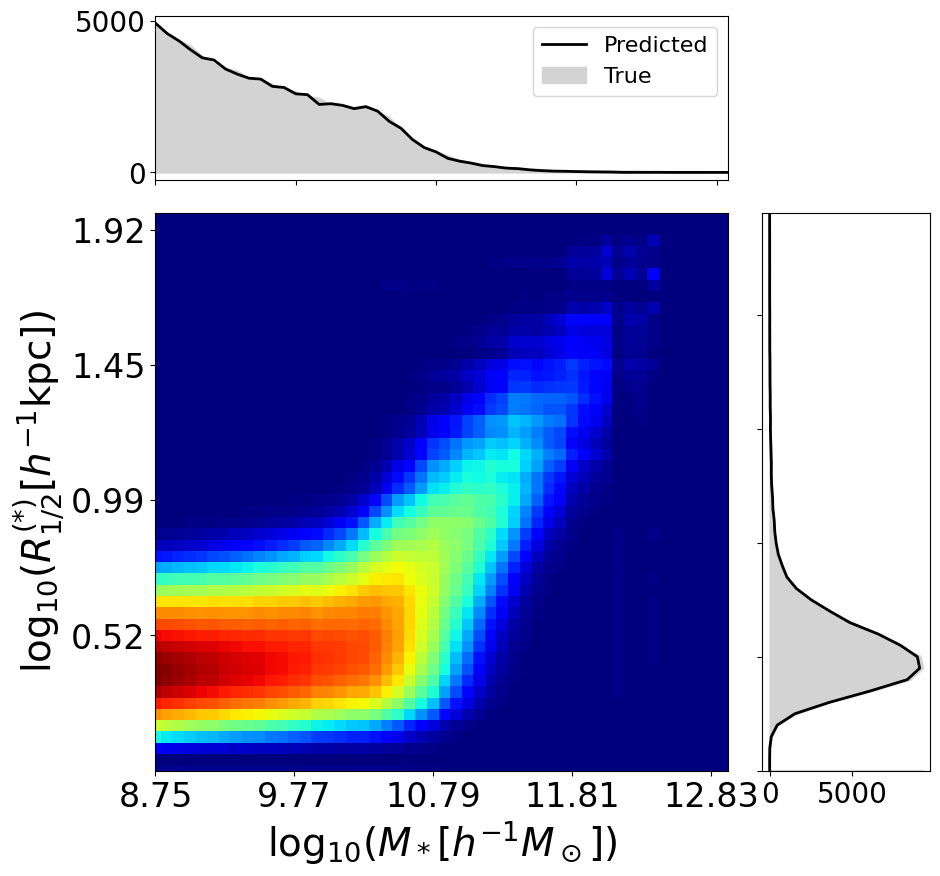}
    \includegraphics[scale=0.225]{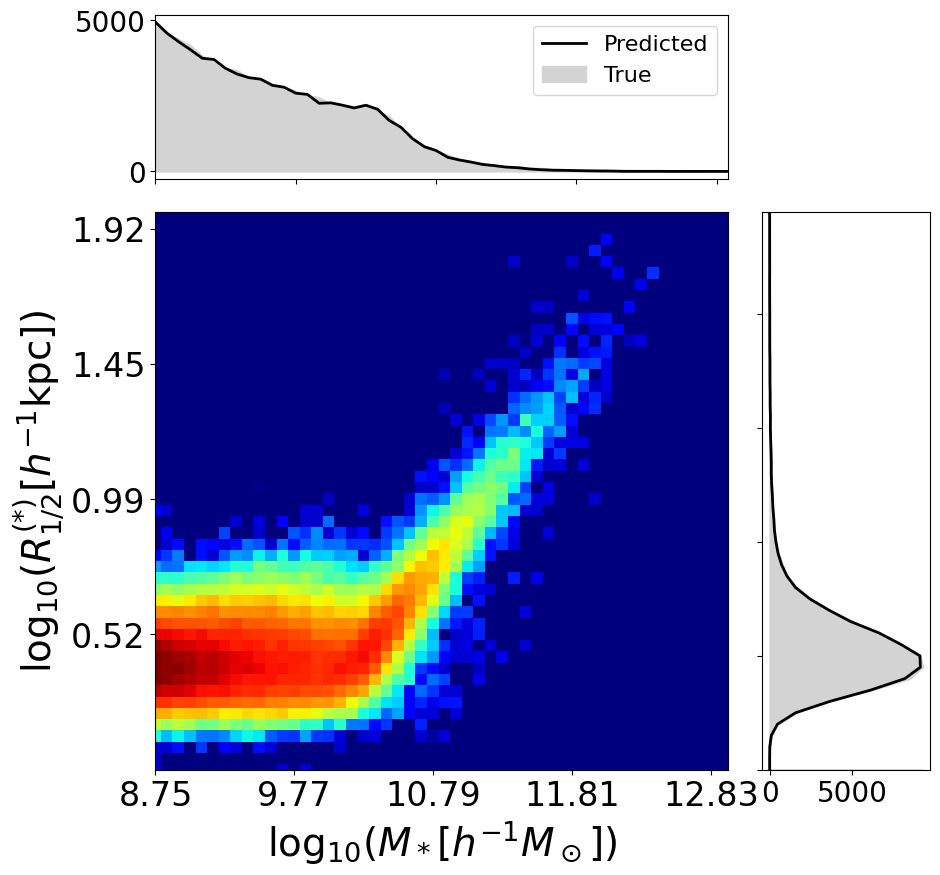}
    \caption{Distributions of galaxy properties. 
    From top to bottom: colour $v.$ stellar mass, sSFR $v.$ stellar mass, sSFR $v.$ colour, and radius $v.$ stellar mass. 
    The first column shows the true distributions from TNG300. 
    The second column shows the distributions computed from the univariate distributions as predicted by $\rm NN_{class}$ -- i.e., predicted independently from each other. 
    The third column shows the joint distributions as predicted by $\rm NN_{class}$. 
    The grey shaded regions in the marginal plots correspond to the TNG300 distributions, while the black solid lines correspond to the $\rm NN_{class}$ predictions. 
    The univariate distributions shown in the third column plots were computed by marginalising the joint distributions.}
    \label{fig:joint pred props}
\end{figure*}

\begin{figure}
    \centering
    \includegraphics[width=0.495\linewidth]{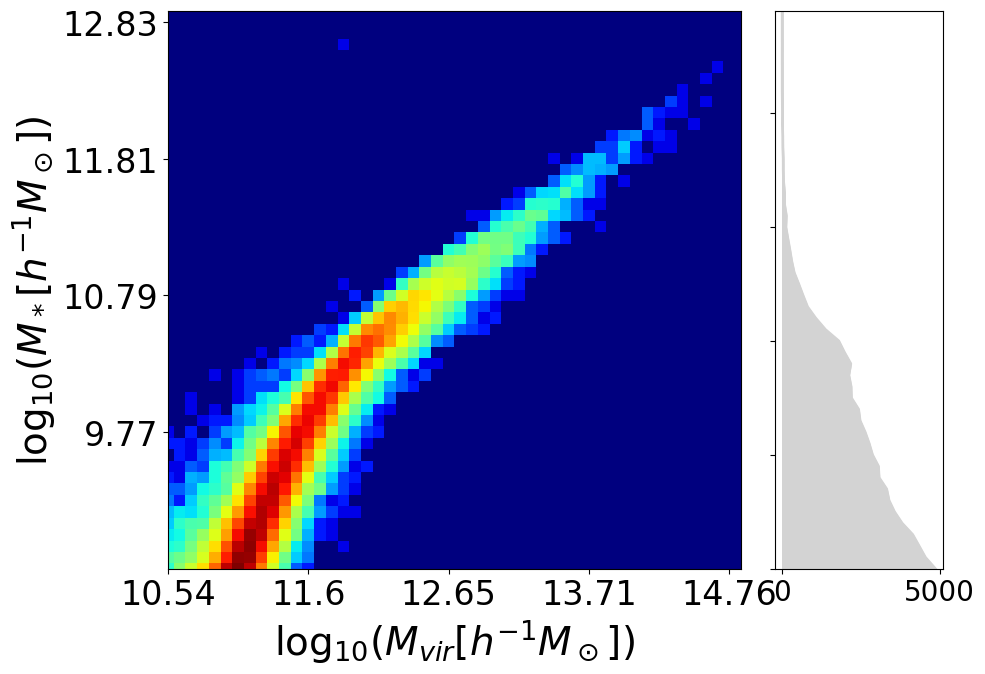}
    \includegraphics[width=0.495\linewidth]{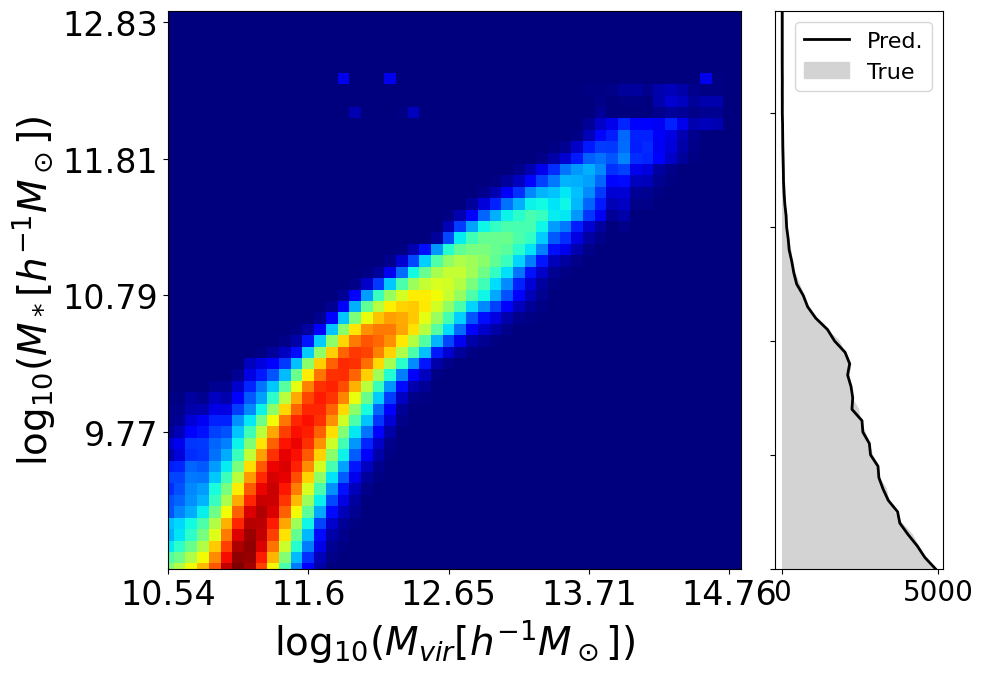}\\
    \includegraphics[width=0.495\linewidth]{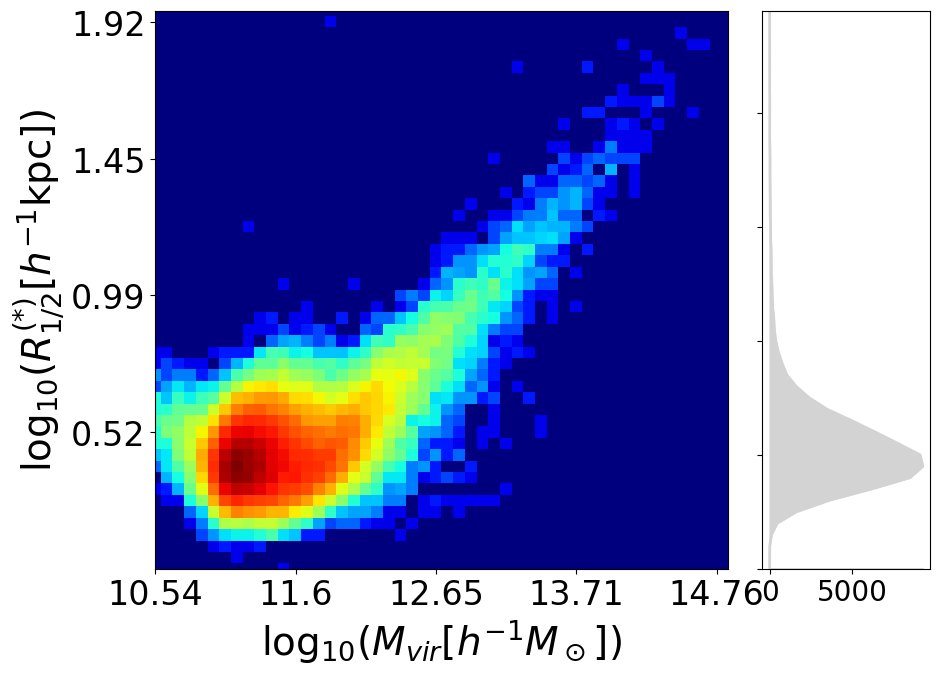}
    \includegraphics[width=0.495\linewidth]{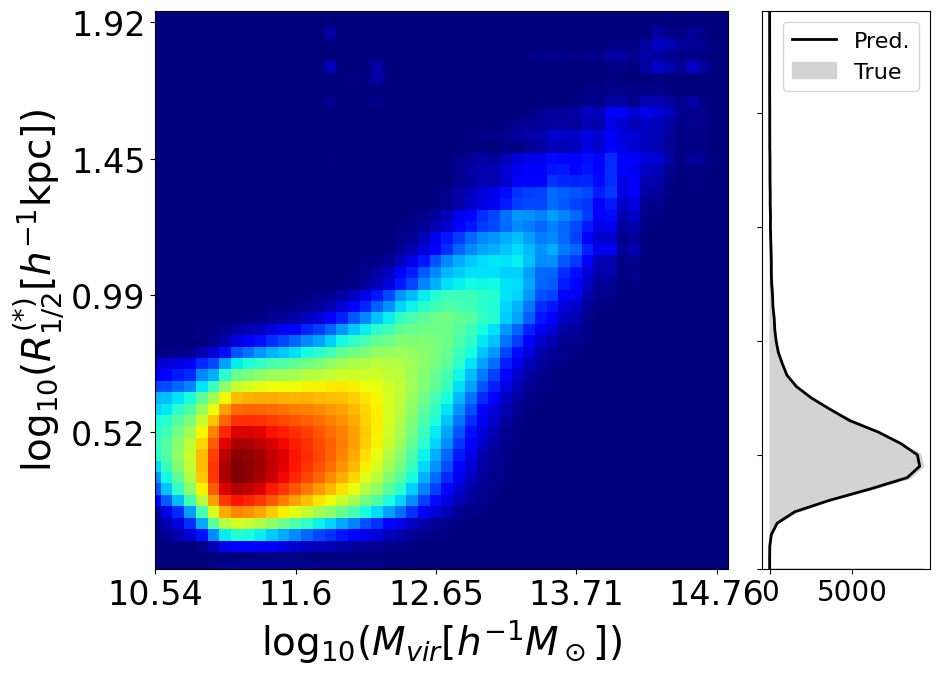}\\
    \caption{Stellar-to-halo mass relation (top) and galaxy size--halo mass relation (bottom) from the TNG300 catalogue (left) and from $\rm NN_{class}$ predictions (right).}
    \label{fig:joint hmass props}
\end{figure}

Fig. \ref{fig:joint pred props} shows the distributions of the galaxies in the test set. 
The first column is the truth table, the TNG300 catalogue. 
The second column is the $\rm NN_{class}$ prediction of univariate distributions, i.e., galaxy properties predicted independently. 
With the univariate distributions we can compute the joint distributions as $P(Y_1) \cdot P(Y_2)$, which are shown in the heatmap diagrams. 
The third column is the $\rm NN_{class}$ prediction for the joint distributions $P(Y_1,Y_2)$, which can be integrated to recover the univariate distributions $P(Y)$ shown in the marginal plots from the third column, i.e.:
\begin{equation}
\label{eq:marg pdf}
    P(Y_i) = \int P(Y_i, Y_j) dY_j.
\end{equation}

The univariate distributions predicted by $\rm NN_{class}$, shown in black solid lines in the second-column plots of Fig. \ref{fig:joint pred props}, are in excellent agreement with the true distributions from TNG300, shown in gray shaded regions. 
They also reproduce fairly well the joint distributions $P(Y_1) \cdot P(Y_2)$ for most cases. The 
$P(g - i) \cdot P(\text{sSFR})$ joint distribution, however, fails to reproduce the shape of the distribution for redder colours and lower sSFRs. 
According to this prediction, red galaxies could have virtually any value of sSFR, while what we actually observe in TNG300 is that as galaxies move from the blue to the red the peak, their sSFRs decrease. 
This important feature is recovered when $\rm NN_{class}$ is trained to predict $P(g-i, \rm sSFR)$ jointly (third column in Fig. \ref{fig:joint pred props}).

The above result indicates that our input halo properties alone are unable to predict accurately the correlations between colour and sSFR. The model would need additional features in order to capture this relation.
It is interesting, however, that we can overcome this limitation by predicting the joint distribution directly using only the presented halo properties.
This exercise indicates that, in order to robustly assign galaxies to haloes, with all the properties consistently correlated, the properties should be predicted together.
Note that, in principle, one could define galaxy populations based on as many parameters as wished. 
Therefore, in the most general case, we would have an $N$-dimensional distribution associated to each host halo.

As a complementary analysis, Fig. \ref{fig:joint hmass props} shows two additional well-known relations in the context of the halo--galaxy connection: the stellar-to-halo mass relation, and the galaxy size--halo mass relation obtained with TNG300 and with $P(M_*)$ and $P(R_{1/2}^{(*)})$ predicted by $\rm NN_{class}$.

Figures \ref{fig:joint pred props} and \ref{fig:joint hmass props} allow for a visual inspection of the results. 
In order to quantify the similarity between the distributions, we have performed the Kolmogorov-Smirnov (KS) test, which measures the maximum distance between cumulative distributions (for more details, see \citealt{ivezic2014statistics}):
\begin{equation}
    \text{KS\ test\ values:}\ \ \Delta = \textit{max} (|F_1 - F_2|).
\end{equation}
The results are shown in Table \ref{tab:KS}. 
For comparison, we also show the values obtained with our baseline models, Raw and SMOGN, from \cite{deSanti2022}. 
%We have used the same grid to build the targets and to compute the cumulative distributions, which may lead to small differences between the numbers presented here and those shown in the aforementioned paper.
Once again, we see that for most cases the independent prediction of univariate distributions reproduce fairly well the joint distributions, except for colour and sSFR.
%$P(g - i, \rm sSFR)$, for which the value becomes considerably lower as compared to $P(g - i) \cdot P(\rm sSFR)$.
In all cases, $\rm NN_{class}$ provides significantly lower values as compared to Raw and SMOGN.

\begin{table*}
\centering
\caption{KS test values for univariate (1D) and joint (2D) distributions computed with the NNs and the baseline models.}
\begin{tabular}{l|c|c|c|l|c|c|c|c}
\hline
\textbf{1D KS} & $P(Y)$ & Raw & SMOGN & \hspace{0.5cm} \textbf{2D KS} & $P(Y_1) \cdot P(Y_2)$ & $P(Y_1 , Y_2)$ & Raw & SMOGN \\
\hline
\vspace{0.1cm}
$P(M_*)$ & 0.002 & 0.064 & 0.064 &\hspace{0.5cm} $P(M_* , g - i)$ & 0.010 & 0.005 & 0.183 & 0.163 \\
\vspace{0.1cm}
$P(g - i)$ & 0.004 & 0.181 & 0.116 &\hspace{0.5cm} $P(M_* , \text{sSFR})$ & 0.012 & 0.009 & 0.253 & 0.209 \\
\vspace{0.1cm}
$P(\text{sSFR})$ & 0.004 & 0.213 & 0.168 &\hspace{0.5cm} $P(g - i , \text{sSFR})$ & 0.110 & 0.009 & 0.266 & 0.176 \\
\vspace{0.1cm}
$P(R^{(*)}_{1/2})$ & 0.009 & 0.217 & 0.110 &\hspace{0.5cm} $P(M_* , R^{(*)}_{1/2})$ & 0.015 & 0.007 & 0.217 & 0.150 \\
\vspace{0.1cm}
 & & & &\hspace{0.5cm} $P(M_\text{vir} , M_*)$ & 0.008 & -- & 0.064 & 0.064 \\
\vspace{0.1cm}
 & & & &\hspace{0.5cm} $P(M_\text{vir} , R^{(*)}_{1/2})$ & 0.012 & -- & 0.217 & 0.110\\
\hline
\end{tabular}
\label{tab:KS}
\end{table*}

\begin{figure*}
    \centering
    \includegraphics[width=0.325\linewidth]{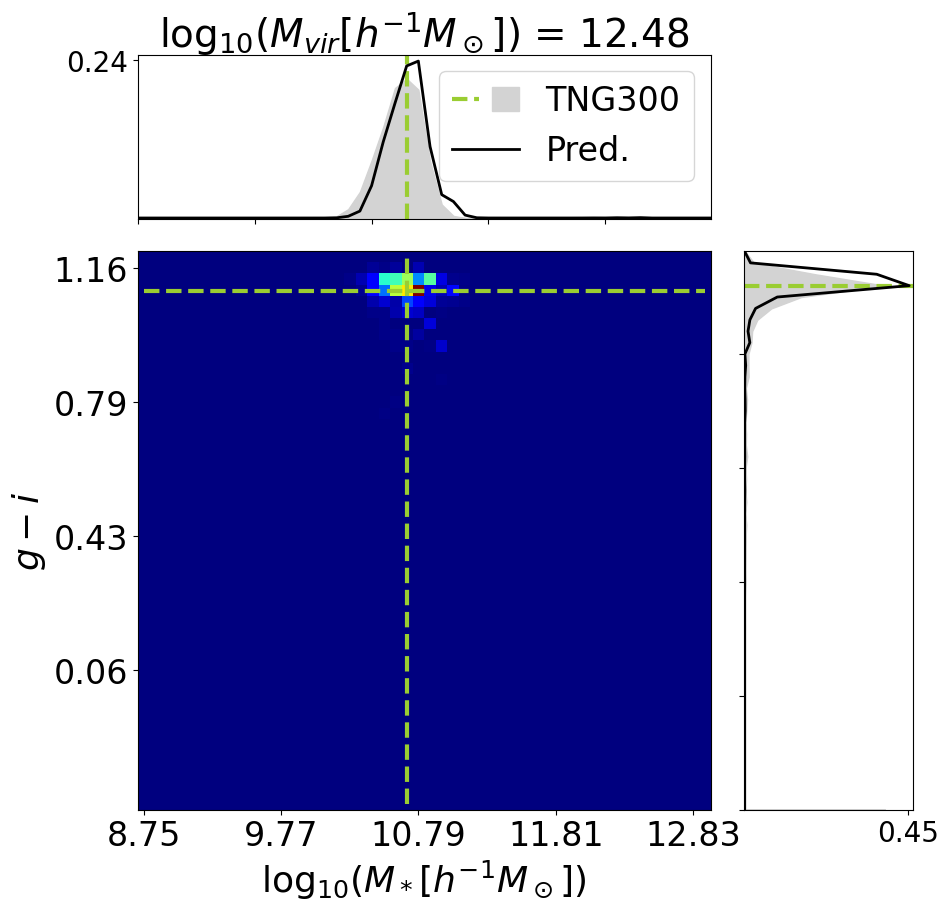}
    \includegraphics[width=0.325\linewidth]{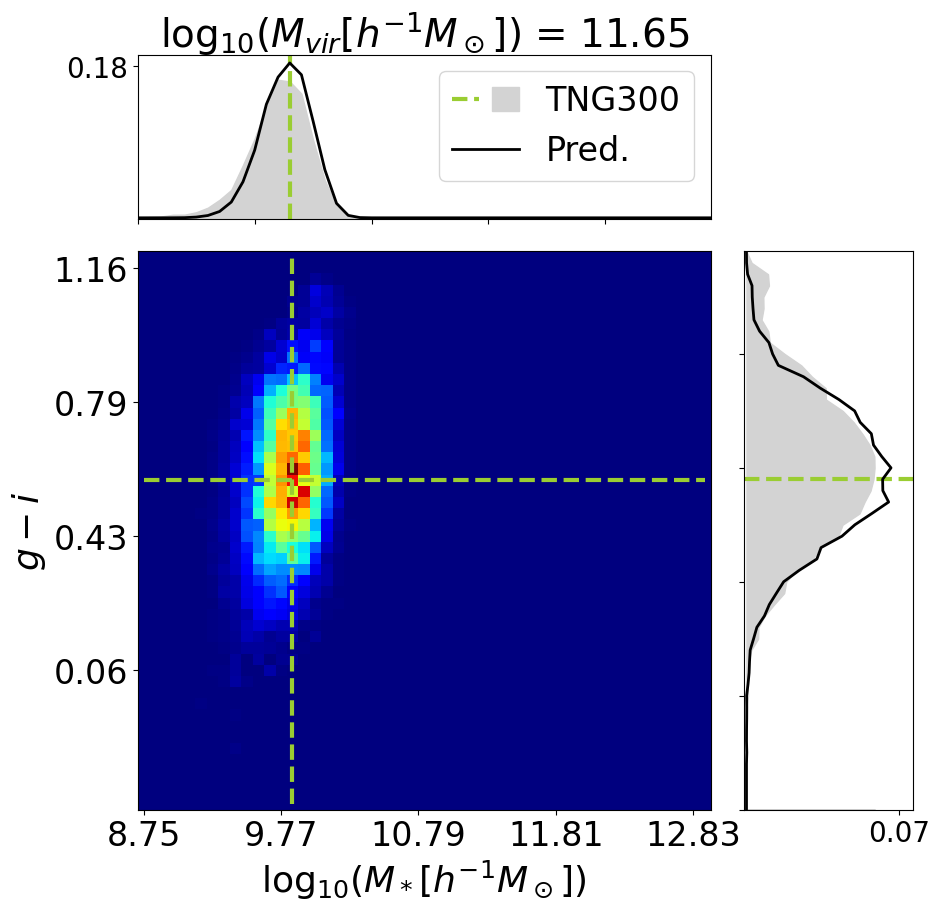}
    \includegraphics[width=0.325\linewidth]{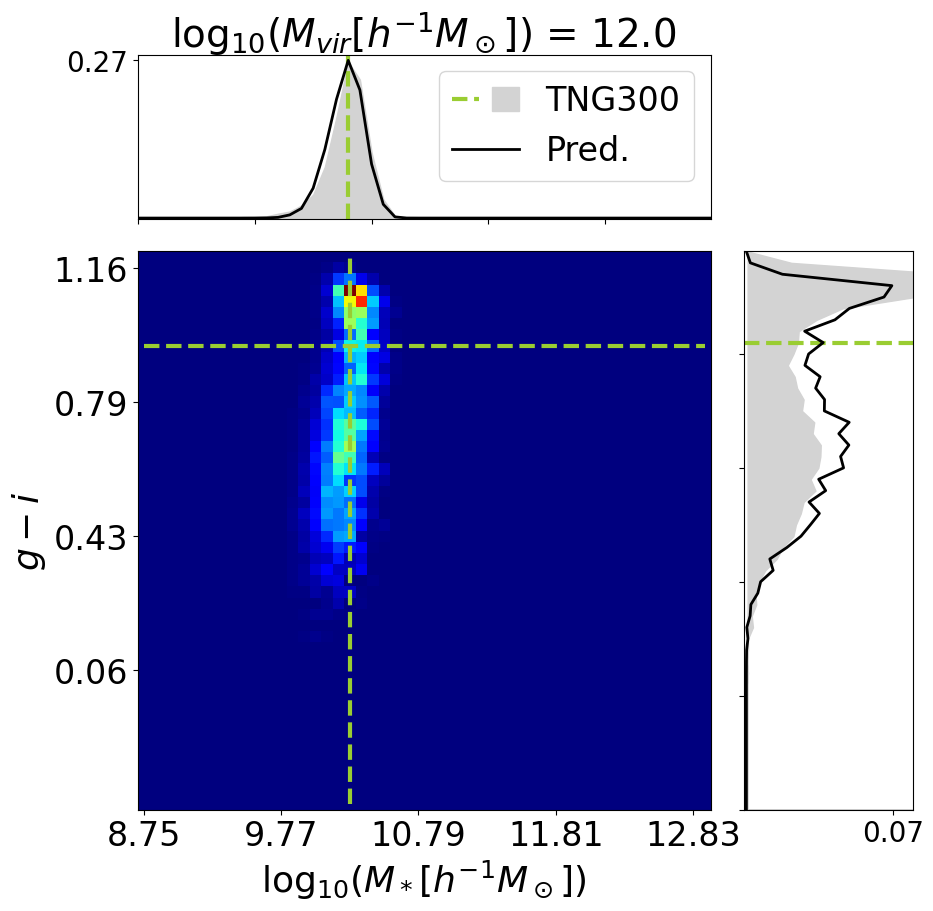}
    \caption{$P(M_*, g - i)$ for individual objects predicted by $\rm NN_{class}$. The dashed green lines show the true values for stellar mass and colour from TNG300. The shaded regions in the marginal plots are the distributions of objects with similar halo mass as indicated on the top of the corresponding panel.}
    \label{fig:ind gals}
\end{figure*}

So far, we have focused on the combined distributions for the entire test sample. 
We now turn our attention to individual objects and the probability distributions that our ML machinery predicts for them. 
In particular, Fig. \ref{fig:ind gals} displays, in a similar format to that of Fig. \ref{fig:joint pred props}, some examples of the joint probability distribution $P(M_*, g-i)$ for three illustrative cases: a red object, a blue object, and an object lying at the so-called green valley region (from left to right). 
In each panel, the host halo mass is specified on the top, whereas the true TNG300 values of stellar mass and colour are shown as the dashed lines.
As a reference, we also include in the marginal plots the distributions of the objects in the test set within a bin of $\pm 0.1$ in halo mass around the values indicated on the top of the plots.

The first thing to notice from Fig. \ref{fig:ind gals} is that the distributions are significantly narrower along the x-axis, as compared to the y-axis. 
This is of course expected, since stellar mass is the galaxy property that displays a tighter relation with the halo properties (particularly with halo mass), and therefore is the easiest to predict. 
It is also noteworthy that not all distributions can be well approximated by a Gaussian distribution. 
Some distributions are significantly skewed or, depending on halo mass, even bimodal, reflecting the well-known colour/sSFR bimodality of the galaxy population (e.g., \citealt{Baldry2004}).

The red galaxy on the left-hand panel shows very little scatter in colour. 
This is typically the case for red galaxies hosted by haloes with $\log_{10} (M_{\rm vir}[h^{-1}\rm M_\odot]) \gtrsim 12.5$. 
By visually inspecting Fig. \ref{fig:joint pred props} and Fig. \ref{fig:joint hmass props}, we can get a sense as to why this happens: massive haloes are typically populated by massive galaxies, since the scatter in the stellar-to-halo mass relation is small. 
Massive galaxies are almost exclusively very red, which explains why the machine predicts a very narrow distribution of colours from the set of halo properties employed. 
The situation is very different for the blue galaxy featured in the middle panel. 
In this case, the predicted colour distribution is much broader than that for the red galaxy. 
Here, the host halo mass is much smaller, which implies a larger scatter in the stellar-to-halo mass relation. 
On top of that, blue galaxies intrinsically display a wide range of colours. 
All this uncertainty is captured by the machine in terms of a wider colour distribution.  

Finally, the green-valley galaxy on the right-hand panel of Fig. \ref{fig:ind gals} represents the most extreme case of the three, where the colour degeneracy produces a bimodal distribution. 
These objects are caught between two intrinsically different populations, i.e., the blue cloud and the red sequence. 
The analysis of individual distributions reveals that these objects are the ones that display a weaker relation with the properties of their host haloes (at least the ones analysed in this work). 
As discussed in \cite{deSanti2022}, these objects exemplify the most clear case where halo properties alone seem insufficient to predict the colour/sSFR, thus emphasising the advantages of our probability-based methodology. 

This probability distribution description on an individual-object basis allows us to explore the dependence of galaxy properties on secondary halo properties at fixed halo mass (a dependence that is closely related to the so-called galaxy assembly bias effect, see, e.g., \citealt{Wechsler2018, SatoPolito2019, MonteroDorta2020_illustris, MonteroDorta2021B}). 
In particular, we have analysed the dependence of $P(M_*, g- i)$ on halo age at fixed halo mass for green-valley objects. 
To this end, we selected objects in the test sample with predicted colour within the range $0.80 < g - i \leq 1.05$ and halo masses of $11.8 < \log_{10}(M_{\rm vir} [h^{-1}\rm M_\odot]) < 12.2$ (we have checked that choosing a narrower halo mass range would not alter our results significantly). 
This subset was subsequently split by halo age (taking the 15$\%$ and 85$\%$ quantiles). 
For younger haloes, a stack of all distributions still reveals some bimodality in colour, albeit with a stronger preference for the blue peak. 
The predicted probability distribution for green-valley galaxies in older haloes is, conversely, much more skewed towards redder colours. 
The tail of the distribution for these objects still covers the green valley, which means that in some realisations these host haloes will be populated by a green-valley central galaxy (although the probability for this to happen is low). 
These results are reassuring in terms of the robustness of our methodology, demonstrating that our probability description is capable of capturing secondary halo dependencies.

\section{Power Spectrum}\label{power spectrum}

With the help of the method presented in this work we have greater flexibility to define different tracers based on galaxy properties. 
In this section, we explore the performance of $\rm NN_\text{class}$ in terms of the accuracy with which we can reproduce the power spectra of those tracers. 
We compute spectra for tracers in the test set, using the python package \texttt{nbodykit} \citep{nbodykit2018}.
For the truth TNG300 catalogue we use the positions of the central galaxies, but for the predictions we use the positions of the host haloes. 
Once again, we compare $\rm NN_{class}$ with the baseline models from \cite{deSanti2022}. 
As a complementary analysis, in Appendix \ref{add pk} we compare the power spectra of tracers defined according to the same criteria of that previous work, which are based on individual galaxy properties.
%, i.e., split only $P(Y)$ instead of $P(Y_1, Y_2)$.

Since TNG300 is a single box, the uncertainties of the spectrum on each bandpower $k_i$, for each tracer $\alpha$, are computed according to the theoretical (Gaussian) covariance, i.e.:
\begin{equation}
\label{eq:sigma pk}
    \frac{\sigma^2_{\alpha, i}}{P^2_{\alpha, i}} = \frac{2}{V\Tilde{V}} \Bigg( \frac{1 + \Bar{n}_\alpha P_{\alpha, i}}{\Bar{n}_\alpha P_{\alpha, i}} \Bigg)^2 \, ,
\end{equation}
with $\Tilde{V} = 4 \pi k_i^2 \Delta k/(2 \pi)^3$, and the residuals are defined as
\begin{equation}
    \frac{\Big(P^\text{pred}_{\alpha, i} - P^\text{TNG300}_{\alpha, i}\Big)^2}{ \sigma^{2}_{\alpha, i} } \, .
\end{equation}

Our choice of tracers is driven by the fact that the target selection in galaxy surveys often rely on the analysis of colour-magnitude diagrams (see e.g. \citealt{Eisenstein_2001, Eisenstein_2011, Zhou_2020}).
One of the most common ways to define galaxy populations is in terms of the red sequence and the blue cloud, which can also be clearly distinguished in the colour-stellar mass diagram, as shown in Fig.\ref{fig:joint pred props}. 
They are two distinct populations with different biases, hence their interest for studies of large scale structure.

\begin{figure*}
    \centering
    \includegraphics[width=1.0\linewidth]{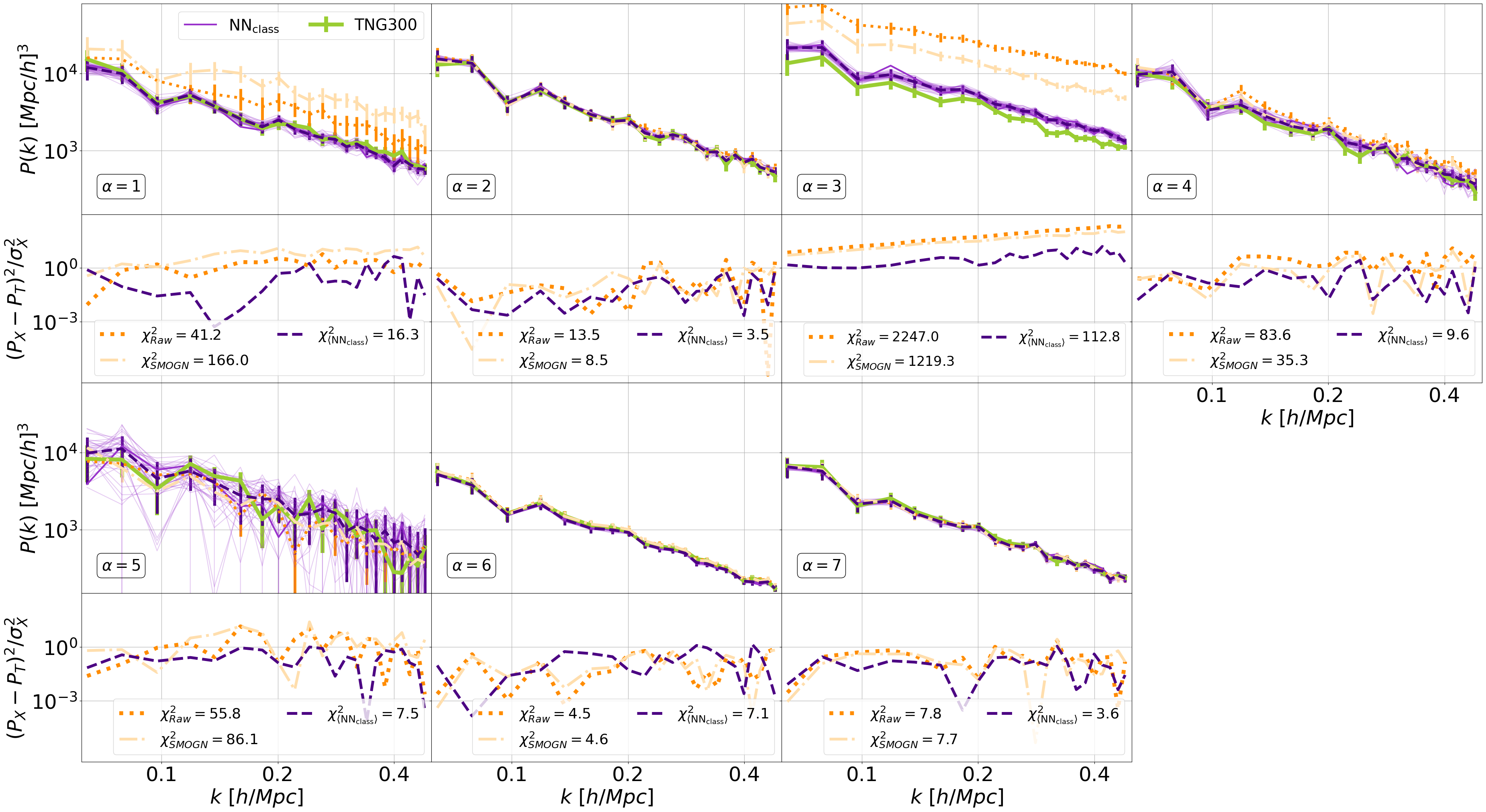}\llap{\includegraphics[height=4.25cm,width=0.24\linewidth]{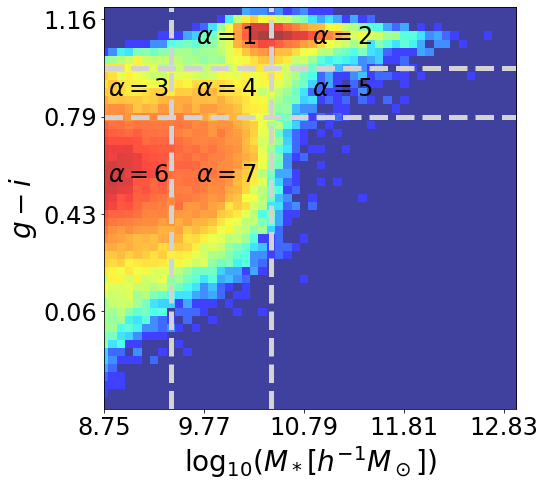}}
    \caption{Power spectra and residuals for seven tracers selected on the basis of the colour-stellar mass diagram (bottom right panel). 
    The green solid lines correspond to TNG300, while the light purple solid lines correspond to spectra from $r=42$ samples drawn from the probabilities predicted by $\rm NN_{class}$. The 
    dark purple, thick dashed lines correspond to the mean of those realisations. 
    The baseline models are shown in orange: darker dotted lines correspond to the Raw model and lighter dotted-dashed lines correspond to the SMOGN model.}
    \label{fig:pk 7 tracers}
\end{figure*}

In a similar fashion, we defined seven tracers ($\alpha = 1, \dots, 7$) based on the colour-stellar mass diagram, $P(M_*, g-i)$.
We split red galaxies ($g - i > 1.05$) into lower ($\alpha=1$) and higher ($\alpha=2$) stellar masses. 
Conversely, ``green-valley'' galaxies (defined as $0.80 < g - i \leq 1.05$) are split into three mass bins, leading to populations $\alpha=3, 4, 5$. 
Finally, blue galaxies ($g - i \leq 0.8$) are separated into lower ($\alpha=6$) and higher ($\alpha=7$) stellar mass bins. 
This selection is outlined in Table \ref{tab:7 tracers}, and it is represented in the lower right corner of Fig. \ref{fig:pk 7 tracers}.

\begin{table}
\centering
\caption{Criteria for splitting central galaxies by stellar mass and colour, in order to define the tracers used in the power spectrum analysis.}
\begin{tabular}{l|c|c|r}
\hline
Tracer & $\log{(M_*[h^{-1}\rm M_\odot])}$ & $g-i$ & \# objects\\
\hline
$\alpha = 1$ & $(9.5, 10.5]$ & $(1.05, )$ & 4,073\\
$\alpha = 2$ & $(10.5, )$ & $(1.05, )$ & 5,207\\
$\alpha = 3$ & $( , 9.5]$ & $(0.80, 1.05]$ & 4,786\\
$\alpha = 4$ & $(9.5, 10.5]$ & $(0.80, 1.05]$ & 5,950\\
$\alpha = 5$ & $(10.5, )$ & $(0.80, 1.05]$ & 1,267\\
$\alpha = 6$ & $(, 9.5]$ & $(, 0.80]$ & 29,695\\
$\alpha = 7$ & $(9.5, 10.5]$ & $(, 0.80]$ & 18,432\\
\hline
\end{tabular}
\label{tab:7 tracers}
\end{table}

An interesting feature of the probabilistic approach is that each galaxy is generated through a realisation of a probability distribution spreading over many bins.
As a consequence, we can build many catalogues of central galaxy properties by drawing values $y_1, y_2$ from $P(Y_1, Y_2)$. 
We have performed $r = 42$ realisations of $P(M_*, g - i)$, leading to as many values of $M_*$ and $g - i$ for each halo. 
We then compute the spectrum of each of these samples, and from that the mean and variance of the spectra. %$\Bar{P}_{\alpha, i} = \sum_{r=1}^{42} P_{\alpha, i}^{r}$.
For the mean spectrum $\Bar{P}_{\alpha, i}$, we compute the uncertainties according to Eq. \eqref{eq:sigma pk}.

Fig. \ref{fig:pk 7 tracers} shows the power spectra and residuals of the seven tracers defined in terms of  $P(M_*, g- i)$ -- see Table \ref{tab:7 tracers}.
Tracers $\alpha = 3, 4$ are relatively rare, hence their corresponding regions in colour-stellar mass space are poorly populated by single-point estimators. 
Therefore, a model that predicts galaxies in these regimes improves the quality of the fit considerably -- i.e., it reduces the $\chi^2$. 
We had already seen an improvement with the SMOGN models, which better recover this region as compared to the Raw models, but with $\rm NN_{class}$ this improvement is even more pronounced.
There are only a few $\alpha = 5$ galaxies in TNG300, which makes this population very sparse. 
In particular, it has the largest variance over realisations.
Conversely, all models are equally good at reproducing the
power spectra of tracer populations closer to the peaks of the probability distributions: for $\alpha=1, 2, 6, 7$, the $\chi^2$ is comparable between all models.

\begin{figure*}
    \centering
    \includegraphics[width=1.0\linewidth]{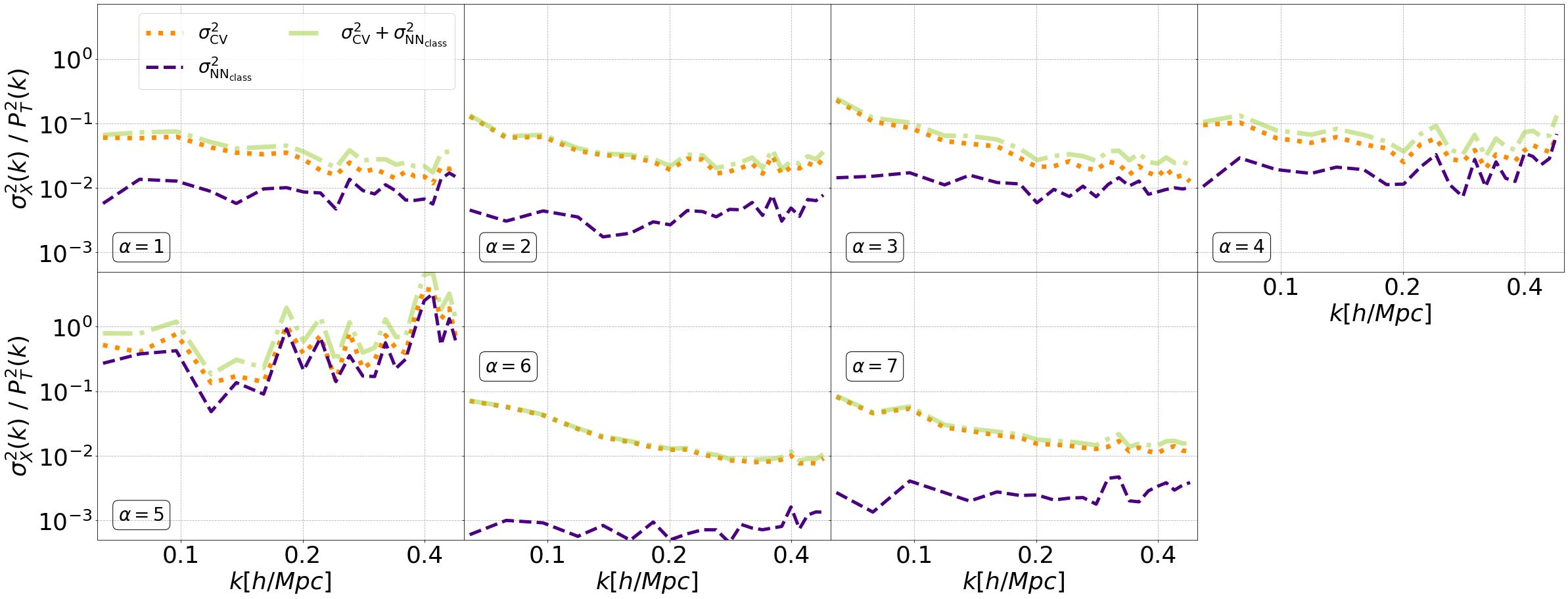}%\llap{\includegraphics[height=3cm,width=0.2\linewidth]{7tracers.png}}
    \caption{Relative error for seven tracers selected based on the colour - stellar mass diagram. The variances are normalised by the TNG300 spectrum $P_T(k)$ of each tracer $\alpha$. Orange dotted lines correspond to the relative error computed with Eq.\eqref{eq:sigma pk}, purple dashed lines correspond to the relative error computed with $\rm NN_{class}$ and green solid lines correspond to the total relative error.}
    \label{fig:error bars pk}
\end{figure*}

As discussed above, we are able to draw multiple samples from the probabilities predicted by $\rm NN_{class}$. 
Each realisation leads to slightly different power spectra, as can be seen in Fig. \ref{fig:pk 7 tracers}.
By computing the variance of the multiple $P(k)$ we can assess the uncertainties due to the intrinsic stochasticity in the halo--galaxy connection.
Fig. \ref{fig:error bars pk} compares the relative errors $\sigma^2/P^2_{\rm TNG300}(k)$ computed using $\sigma^2_{\rm CV}$, from Eq. \ref{eq:sigma pk} (which encodes the uncertainty due to cosmic variance, CV), with $\sigma^2_{\rm NN_{class}}$, which encodes the statistical uncertainties in the halo--galaxy connection estimated with $\rm NN_{class}$.
As we already saw in Fig. \ref{fig:pk 7 tracers}, the cosmic variance error bars are typically larger than the scatter in the power spectra due to the multiple realisations of the $\rm NN_{class}$ probabilities. 
The contribution of $\sigma^2_{\rm NN_class}$ seems more relevant for the tracer population 5, which is very sparse.
However, for all tracers $\sigma_{\rm CV}^2$ decreases for smaller scales (due to the Fourier bin volume), while $\sigma_{\rm NN_{class}}$ remains approximately constant.
Therefore, the relative contribution of $\sigma_{\rm NN_{class}}$ for the total error budget of the power spectra appears to become more important at smaller scales.

Even though we see no evidence of a bias associated with this additional source of statistical uncertainties, the stochastic nature of the relationship between galaxies and their haloes may present further challenges for multi-tracer analyses of LSS \citep{Seljak:2008xr,McDonald:2008sh}. 
The advantages of the multi-tracer technique are reliant upon the partial cancellation of cosmic variance that results from clustering measurements from different galaxy types that are assumed to reflect the same underlying dark matter density field -- in that respect see also \citet{Abramo2013,MTPK}.
The ``stochastic bias'' associated with the nature of the galaxy--halo connection can dilute some of the expected cosmic variance cancellation.
However, that stochastic component seems to affect mostly the power spectra on small scales, where non-linear effects already limit our ability to employ the multi-tracer technique effectively -- see, e.g., \citet{MonteroDorta2020_MTPK}.

\section{Discussion and Conclusions}\label{conclusions}

Although there is an obvious relation between the baryonic and DM components of haloes, there is also mounting evidence that the properties of haloes alone are insufficient to reproduce the properties of galaxies, since the latter are shaped by a variety of galaxy-formation processes. On the other hand, ML regression models are traditionally designed to reproduce single-value statistics, and thus are ill-equipped to encode the intrinsic scatter in the halo--galaxy connection. Building on the recent work of \cite{deSanti2022}, here we use the TNG300 hydrodynamical simulation in combination with NNs to map the connection between the properties of central galaxies and the properties of their hosting haloes. As in the aforementioned work, NNs are trained to reproduce the stellar mass, $g - i$ colour, sSFR and radius of TNG300 galaxies based on a set of halo/environmental properties that include virial mass, concentration, formation redshift, spin, and overdensity (computed over scales of 3 $h^{-1}$Mpc). In order to alleviate the deficiencies of ML deterministic regression models, we have tested a different approach for the first time in the context of the halo--galaxy connection. The NNs are now trained to predict probability distributions instead of single-value statistics by means of a binning classification scheme. 
In essence, the distributions of galaxy properties are split into $K$ narrow bins so that the NNs can associate a score to each of the $K$ classes. This is performed in such a way that the output can be used as a proxy for the probability distributions of the central galaxy properties.

We have shown that this approach is in fact capable of producing bivariate distributions of galaxy properties, i.e., $P(Y_1, Y_2)$, in outstanding agreement with those from TNG300 (here, $\lbrace Y_1,Y_2 \rbrace$ is any pair of galaxy properties). These joint distributions can be compared with the product of the two 1D (disjoint) distributions, $P(Y_1)$ and $P(Y_2)$.
For the joint distributions, we employ 2D $K \times K$ grids, representing the binned galaxy properties, where each pixel on the grid corresponds to a class. In either case, predicting the probability distributions yields significantly better results compared with the deterministic approach \citep{deSanti2022}, as both a visual inspection and the 2D KS test reveal. As a reference, our 2D KS test for the joint distributions $P(Y_1, Y_2)$ yields performance results that are better by factors of 10-30 as compared to those reported in \cite{deSanti2022}. We have also checked that predicting galaxy pairs directly is particularly advantageous for the colour--sSFR joint distribution, where the stellar mass, the main anchor of the halo--galaxy connection, is not included. 

An important sub-product of our analysis are the joint distributions for individual galaxies, which can be understood as the probability distributions that an object occupies a given location on the 2D diagrams for the galaxy properties. As an illustration, we have analysed the individual joint distributions of stellar mass and colour, and verified that the distributions for red galaxies, particularly for those that live in massive haloes, are significantly more concentrated than those for blue and green-valley objects. 
For the latter, the individual distributions can even become bimodal in certain halo mass ranges. This is a robustness test for our methodology, showing that these individual distributions are good estimators of the uncertainty that results from attempting to predict galaxy properties from incomplete (halo) information. The main advantages of our method are that it provides a more complete description of the interconnected relations between galaxy and halo properties, as compared to single-value ML approaches, and that it can be easily implemented in cosmological and galaxy formation models.

As an application of our methodology, we have shown that our predictions are capable of reproducing with unprecedented precision the power spectra of any given number of tracers defined based on the colour-stellar mass diagram (we showed results for 7 tracers, but the analysis can be extended to more galaxy populations). 
We have also checked that the statistical uncertainty in our models (which can be obtained by sampling the distributions several times, creating multiple catalogues) is often small compared with the uncertainty that emanates from cosmic variance (particularly on large scales). 
In this sense, our method is clearly advantageous for cosmological studies employing a high number of tracers and/or underrepresented populations, as compared with the more traditional single values approaches (see \citealt{deSanti2022} for comparison). 
These advantages can be exploited in the context of multi-tracer cosmological analyses, where clustering information from multiple galaxy population and redshift ranges is combined in order to reduce the uncertainties in the estimation of the power spectrum, and thus the bias and cosmological parameters (e.g., \citealt{Abramo2013, MTPK,MonteroDorta2020_MTPK, Abramo_2022}).

One interesting application of our method is to paint galaxies onto haloes in DM only simulations. 
As we have discussed in this work, when central galaxy properties are predicted jointly, their correlations are in agreement with those from hydrodynamical simulations. However, in order to extend our analysis to a higher number of dimensions, i.e., to predict joint distributions of 3 or more properties, or to extend the approach to satellite galaxies, it is necessary to optimise the discretisation of the galaxy distributions. Presently, our method can become computationally inefficient for this purpose, as so far we are considering bins of equal size across the galaxy property diagrams. Follow-up work will be devoted to improving this methodology in order to generalise the analysis.

Finally, the flexibility of our method in terms of reproducing both the clustering and internal properties of virtually any galaxy population with precision may have applications in the context of galaxy assembly bias, i.e., the secondary dependencies of galaxy clustering at fixed halo mass (see, e.g., \citealt{Lin2016,Zu2016,MonteroDorta2017,Niemiec2018,Zentner2019,MonteroDorta2020_illustris,Obuljen2020,Salcedo2022,Wang2022}). 
In particular, recent attempts to probe the effect with observations \citep{Salcedo2022,Wang2022} have employed forward-modelling techniques using specifically generated galaxy mocks. 
Our methodology and statistical descriptions seem ideal to be incorporated into these models.

\section*{Acknowledgements}
NVNR, NSMS and LRA acknowledge Coordena\c{c}\~ao de Aperfei\c{c}oamento de Pessoal de N\'ivel Superior (CAPES), Funda\c{c}\~ao de Amparo \`a Pesquisa do Estado de S\~ao Paulo (FAPESP), and Conselho Nacional de Desenvolvimento Cient\'ifico e Tecnol\'ogico (CNPq) for financial support. NSMS acknowledges financial support from FAPESP, grant \href{https://bv.fapesp.br/pt/bolsas/187647/matrizes-de-covariancia-cosmologicas-e-metodos-de-machine-learning/}{2019/13108-0} and \href{https://bv.fapesp.br/en/bolsas/202438/machine-learning-methods-for-extracting-cosmological-information/}{2022/03589-4}. ADMD thanks Fondecyt for financial support through the Fondecyt Regular 2021 grant 1210612.

%%%%%%%%%%%%%%%%%%%%%%%%%%%%%%%%%%%%%%%%%%%%%%%%%%
\section*{Data Availability}

The material presented in this paper is available in the repository:
\href{https://github.com/nvillanova/central-galaxies-joint-distributions}{https://github.com/nvillanova/central-galaxies-joint-distributions}.

%The inclusion of a Data Availability Statement is a requirement for articles published in MNRAS. Data Availability Statements provide a standardised format for readers to understand the availability of data underlying the research results described in the article. The statement may refer to original data generated in the course of the study or to third-party data analysed in the article. The statement should describe and provide means of access, where possible, by linking to the data or providing the required accession numbers for the relevant databases or DOIs.



\bibliographystyle{mnras}
\bibliography{main} % if your bibtex file is called example.bib

%\begin{thebibliography}{99}
%\end{thebibliography}

%%%%%%%%%%%%%%%%%%%%%%%%%%%%%%%%%%%%%%%%%%%%%%%%%%

%%%%%%%%%%%%%%%%% APPENDICES %%%%%%%%%%%%%%%%%%%%%

\clearpage

\appendix

\section{Single value estimation}\label{single point}

\begin{figure}
    \centering
    \includegraphics[width=0.75\linewidth]{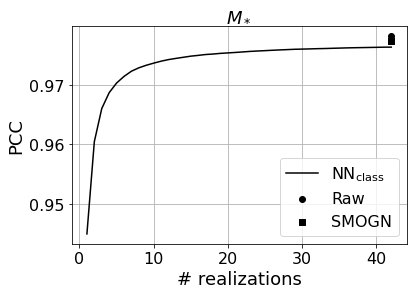}
    \includegraphics[width=0.75\linewidth]{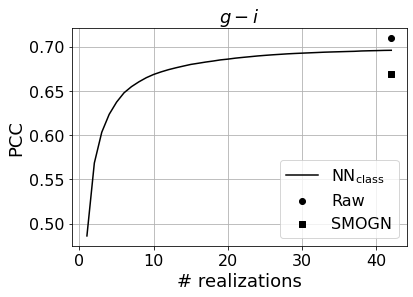}
    \includegraphics[width=0.75\linewidth]{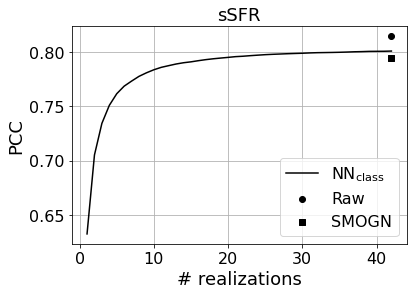}
    \includegraphics[width=0.75\linewidth]{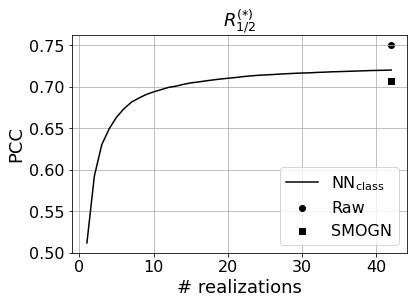}
    \caption{PCC of $\rm NN_{class}$ (solid lines) as a function of the number of realisations of $P(Y), Y = M_*, g-i$, sSFR, $R_{1/2}^{(*)}$. The PCC values of the baseline models Raw and SMOGN are shown as dotted and squared markers, respectively.}
    \label{fig:pcc vs realizations}
\end{figure}

In this appendix we discuss the results of the $\rm NN_{class}$ in terms of single-point estimation scores. 
Throughout the paper, our analysis focus on the performance in terms of how well we can recover the distributions. 
Since we do not have a single value associated to each data set instance, but a distribution, one can sample several times from this distribution in order to estimate the most probable value, and compute single-point estimation metrics with it.
Once again, we take the average of $r=42$ realisations of each predicted galaxy property and calculate the Pearson Correlation Coefficient (PCC) between the true and estimated values as:
\begin{equation}
\label{eq:pcc}
    \rm PCC  =\frac{cov(y^{pred}, y^{true})}{\sigma_{y^{pred}}\sigma_{y^{true}}}.
\end{equation}
Fig. \ref{fig:pcc vs realizations} shows the PCC score as a function of the number of realisations and also the values of the baseline models for the four galaxy properties. In this exercise, we sample from univariate distributions $P(Y)$ instead of joint distributions.
$\rm NN_{class}$ provides results comparable to the single-point estimators Raw and SMOGN as the number of realisations increases, which indicates that $\rm NN_{class}$ are also good maximum likelihood estimators.

\section{Power Spectrum: additional results}\label{add pk}

\begin{figure*}
    \centering
    \includegraphics[width=1.0\linewidth]{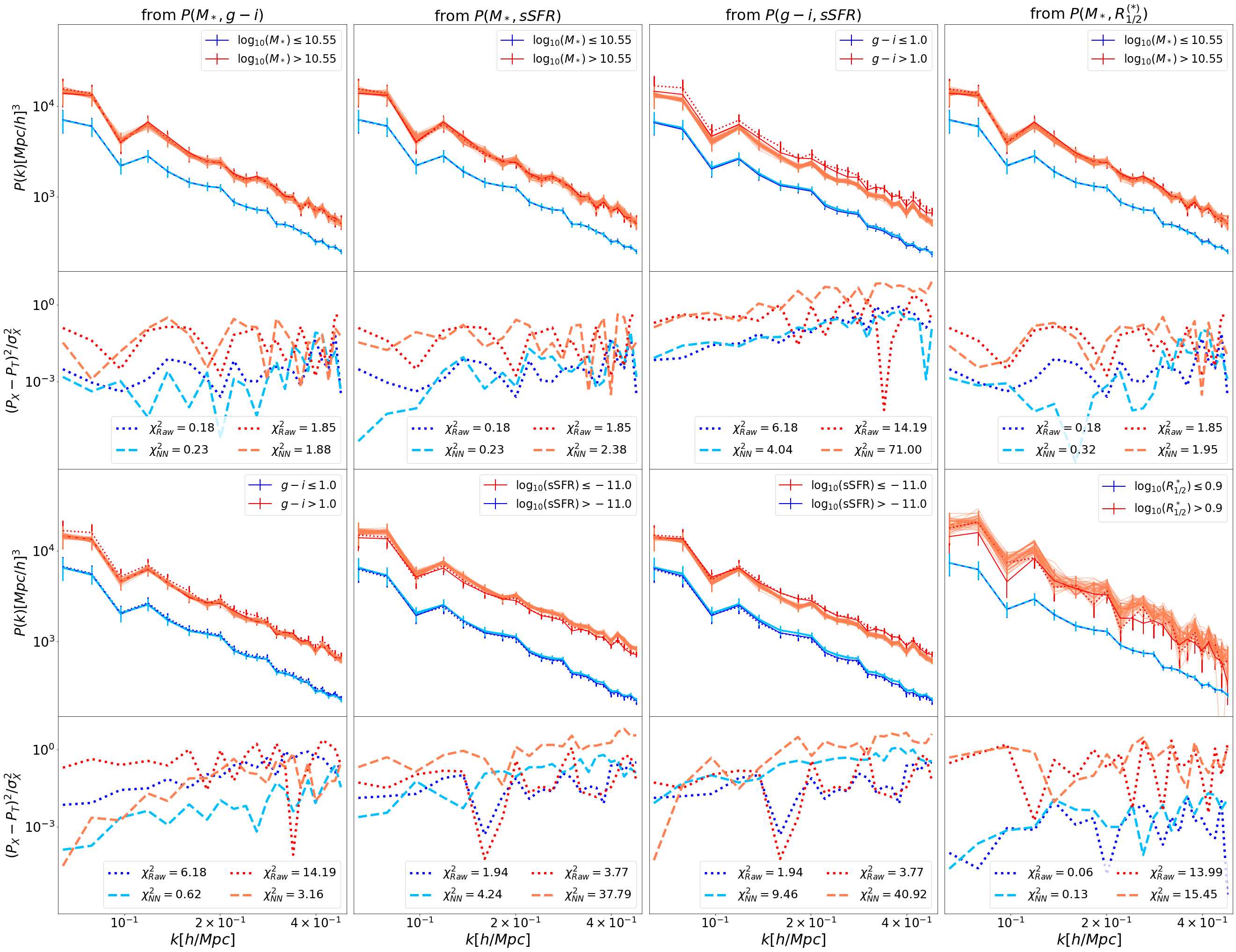}
    \caption{Power spectrum and residuals of two tracers defined by splitting each galaxy property. The higher bias tracers are shown in red, and the lower bias tracers are shown in blue. The properties are obtained by marginalising the joint distributions and can thus be obtained with more than one distribution. The first column shows the results for stellar mass and colour obtained with $P(M_*, g-i)$. The second column shows the results for stellar mass and sSFR obtained with $P(M_*, \rm sSFR)$. The third column shows the results for colour and sSFR obtained with $P(g - i, \rm sSFR)$. The fourth column shows the results for stellar mass and radius obtained with $P(M_*, R_{1/2}^{(*)})$. The power spectrum of each $\rm NN_{class}$ realisation is shown as solid lines. The mean $\rm NN_{class}$ spectra are shown as dashed lines and the Raw model spectra are shown as dotted lines.}
    \label{fig:pk deSanti}
\end{figure*}

In this appendix we show the power spectrum of the tracers defined in \cite{deSanti2022} -- see Fig. \ref{fig:pk deSanti}. The galaxies are divided into two populations based on each of the properties. The univariate distributions can be obtained from different joint distributions, by marginalising them -- see Eq. \eqref{eq:marg pdf}. Stellar mass can be obtained from $P(M_*, g-i), P(M_*, \rm sSFR)$ and $P(M_*, R_{1/2}^{(*)})$, colour can be obtained from $P(M_*, g-i)$ and $P(g - i, \rm sSFR)$, sSFR can be obtained from $P(g - i, \rm sSFR)$ and $P(M_*, \rm sSFR)$, and the radius can be obtained from $P(M_*, R_{1/2}^{(*)})$. Once again, for $\rm NN_{class}$ we show $r=42$ realisations as well as the mean of the spectra of all $r$ samples.
We see that for these tracers there is no clear advantage of the $\rm NN_{class}$ over the Raw model:
in most cases $\rm NN_{class}$ performs similar to the RAW models, although for sSFR the results for $\rm NN_{class}$ are slightly worse (which is not entirely unexpected, since sSFR is a particularly difficult property to predict based only on the halo properties that we take into account).
Note that here we are computing the average of the spectra of many realisations of the predicted distributions, as in Fig. \ref{fig:pk 7 tracers}. 
In this way we can explore the advantage of having a tool that recovers the complete range of possible values. In order to have a more straightforward comparison with the single-point estimators, one can compute the spectrum of the tracers defined based on the maximum likelihood values of galaxy properties, as in Appendix \ref{single point}.

% \textcolor{blue}{[Raul: There is one thing I don't really understand about this last figure. 
% From Fig. 1, the 1D PDFs obtained by marginalising the 2D (joint) PDFs are in perfect agreement with the original 1D ones. 
% So, for this test the two ways of splitting the populations in 1D should be roughly equivalent -- or at least the NNs should perform a bit better, no?... I don't really understand why in some cases (especially for sSFR) the NNs perform worse than the RAW...
% ]}

%{\color{violet} I suspect that the power spectrum performance is somehow a balance between the good prediction of maximum likelihood values and the recovery of the distributions.}

%%%%%%%%%%%%%%%%%%%%%%%%%%%%%%%%%%%%%%%%%%%%%%%%%%

% Don't change these lines
\bsp	% typesetting comment
\label{lastpage}
\end{document}